\documentclass{article}[12pt]
\usepackage{setspace}
\usepackage{comment}
\usepackage{arxiv}

\usepackage[utf8]{inputenc} % allow utf-8 input
\usepackage[T1]{fontenc}    % use 8-bit T1 fonts
\usepackage{hyperref}

\usepackage{url}            % simple URL typesetting
\usepackage{booktabs}       % professional-quality tables
\usepackage{amsfonts}       % blackboard math symbols
\usepackage{nicefrac}       % compact symbols for 1/2, etc.
\usepackage{microtype}      % microtypography
\usepackage{lipsum}		% Can be removed after putting your text content

\usepackage{amsmath}
\usepackage{xcolor}
\usepackage{amsthm}
\usepackage{pdflscape}
\usepackage{pgfplots}
\usepackage{mathrsfs}
\usepackage{mathtools}
\usepackage{algorithmic}
\usepackage{algorithm}

\title{Easy conditioning far beyond Gaussian}

\newtheorem{thm}{Theorem}[section]

\newtheorem{remark}[thm]{Remark}

\title{Easy Conditioning far beyond Gaussian}

\date{} 					% Or removing it

\author{ {Antoine ~Faul}\thanks{corresponding author} \\
	Institute of Mathematical Statistics and Actuarial Sciences\\
	University of Bern, Switzerland\\
	\texttt{antoine.faul@unibe.ch} \\
	\And
	{David ~Ginsbourger} \\
	Institute of Mathematical Statistics and Actuarial Sciences\\
	University of Bern, Switzerland\\
	\texttt{david.ginsbourger@unibe.ch} \\
 \And
	{Ben  ~Spycher} \\
	Institute of Social and Preventive Medicine\\
	University of Bern, Switzerland\\
	\texttt{ben.spycher@unibe.ch} \\
}

\renewcommand{\headeright}{Faul \& al.}
\renewcommand{\shorttitle}{Easy Conditioning far beyond Gaussian}

\hypersetup{
pdftitle={A template for the arxiv style},
pdfsubject={q-bio.NC, q-bio.QM},
pdfauthor={David S.~Hippocampus, Elias D.~Striatum},
pdfkeywords={First keyword, Second keyword, More},
}

\theoremstyle{plain}

\newtheorem{theorem}{Theorem}[section]

\newtheorem{proposition}{Proposition}
\theoremstyle{definition}
\newtheorem{definition}[theorem]{Definition}

\theoremstyle{remark}

\begin{document}
\maketitle

\begin{abstract}

Multivariate Gaussian distributions enjoy Gaussian conditional distributions that makes conditioning easy: conditioning boils down to implementing analytical formulae for conditional means and covariances. For more general distributions, however, conditional distributions may not be available in analytical form and require demanding and approximate numerical approaches.   
Primarily motivated by probabilistic imputation problems, we review and discuss families of multivariate distributions that do enjoy analytical conditioning, also providing a few counter-examples. Proving that trans-dimensional stability under conditioning extends to mixtures and increasing transformations, we demonstrate that a broader class of multivariate distributions inherit easy conditioning properties. Building on this insight, we developed a generative method to estimate conditional distributions from data by first fitting a flexible joint distribution using copulas and then performing analytical conditioning in a latent space. In our applications, we specifically opt for Gaussian Mixture Copula Models (GMCM), comparing in turn various fitting strategies. Through simulations and real-world data experiments, we demonstrate that our method achieves competitive results with state-of-the-art models in conditional density estimation and data imputation, while offering several advantages due to its analytical form and generative nature. We also touch upon links to Gaussian process modelling and how stability by mixtures and transformations carries over towards easy conditioning of non-Gaussian processes.

\end{abstract}
\section{Introduction}\label{sec:intro}

Conditioning multivariate distributions is a crucial yet challenging task in statistics. While conditional densities are easily obtained in the Gaussian case, more flexible multivariate distributions often require computationally intensive methods to obtain approximate conditional samples. In contrast, fast (approximate) evaluations of conditional probability density functions and cumulative distributions functions may be advantageous for computing quantities of interest, such as conditional quantiles and dependence measures. In particular, having explicit probabilistic models in closed or nearly closed form (possibly up to numerically evaluated marginal transformations) has desirable properties of coherence across conditioning queries, downstream mathematical tractability, and rare event modelling. 

Estimating conditional distributions arises in a wide range of statistical applications, including probabilistic forecasting \cite{gneiting2014probabilistic}, the construction of prediction intervals for time series \cite{hall1999methods}, and the probabilistic imputation of systematically missing values \cite{muhlemann2025comparing}. Such methods have been applied across diverse scientific domains, including astrophysics \cite{dalmasso2020conditional}, econometrics \cite{hansen1994autoregressive}, and meteorology \cite{jeon2012using}.

A broad range of approaches has been proposed for this task. Non-parametric methods, such as conditional kernel density estimation \cite{bashtannyk2001bandwidth, de2003conditional}, provide flexible estimates but can become challenging in high-dimensional settings \cite{holmes2012fast} and typically require careful bandwidth selection. Related non-parametric approaches have also been developed for estimating conditional copulas, with established asymptotic consistency properties \cite{gijbels2011conditional}. Parametric alternatives include least-squares density ratio estimation \cite{sugiyama2010conditional} and conditional transformation models \cite{hothorn2014conditional}. More recently, machine learning approaches have demonstrated strong performance in conditional distribution estimation, ranging from distributional random forests \cite{cevid2022distributional} to mixture density networks \cite{bishop1994mixture}.

%\textcolor{red}{While Eaton \cite{eaton1989group} studied the invariance of families of probability distributions under group actions, the corresponding notion of stability conditioning of multivariate distributions has, to the best of our knowledge, not been explicitly defined in the literature. Similar to conditionally specified distributions \cite{arnold2012conditionally}, this stability property would enable the specification of all conditional distributions.  }

In this paper, we explore non-Gaussian families of multivariate distributions that allow analytical computation of conditional distributions such as multivariate Student $t$-distribution \cite{ding2016conditional} or multivariate skew normal distributions \cite{azzalini1996multivariate}. Furthermore, by introducing the concept of trans-dimensional families of probability distributions that remain stable under conditioning, and showing that this principle extends to mixtures and increasing transformations, we demonstrate that a broader class of distributions can benefit from analytical conditioning. In short, analytical conditioning of multivariate densities can be extended far beyond the Gaussian case. Additionally, since the distributions of stochastic processes are characterized by finite-dimensional distributions, we demonstrate how easy conditioning can similarly be extended beyond Gaussian processes (GPs) via increasing transformations and mixtures. 
In particular, if a stochastic process is easy to condition, any increasing transformation of this process will retain this property. This is exemplified by Warped GPs introduced by Snelson \cite{snelson2003warped}. Specifically, Copula processes defined in Wilson and al. \cite{wilson2010copula} inherit the conditioning properties of their base measure. Moreover, our results show that analytical conditioning properties are available for mixtures of known stochastic processes, such as GPs, Student's t-processes \cite{shah2014student}, and skew-normal processes \cite{benavoli2020skew}.

%This insight leads to a flexible generative method for estimating conditional distributions from data. 
Of particular interest here is the problem of approaching the conditional distribution of a random vector $\mathbf{X}_{1} \in \mathbb{R}^{q}$ given an observation $\mathbf{x}_{2}$ of $\mathbf{X}_{2} \in \mathbb{R}^{p}$, based on a previously acquired set of $n$ observations $(\mathbf{x}_{1i},\mathbf{x}_{2i})_{i=1}^{n}$ assumed to have been independently generated from $(\mathbf{X}_{1},\mathbf{X}_{2})$'s (joint) distribution. 
%This setting is commonly encountered in various statistical problems such as imputation of systematically missing values \cite{anja2024}

Assuming conditions of existence to be fulfilled, a natural approach to estimate the conditional density $f(\mathbf{x_{1}}|\mathbf{x_{2}})$ is to first estimate the joint density by $\hat{f}(\mathbf{x_{1}},\mathbf{x_{2}})$ and then perform conditioning. Using our theoretical results, we show that this strategy can be applied far beyond the Gaussian case, which is often limiting due to its inability to capture complex dependencies and asymmetries typical in observed data. %Notably, our approach allows modeling marginal distributions and dependence structures separately, using copulas.
Using copulas, we can model marginal distributions and dependence structures separately. The first step of the approach consists in fitting $\hat{f}$ via marginal and copula modeling. What is special here is that we resort to copula families possessing properties making them easy to condition. 
%which can be challenging especially in high-dimensions.
 In a second step, when conditioning, the values of the conditioning variables are transformed and easy conditioning is performed in a latent space. Finally, resulting conditional distributions or values sampled from them are converted back to the original space.%Typically, these transformations involve evaluating the quantile function, a task that often relies on the bisection algorithm and can be challenging. In this work, we propose a solution to circumvent this issue specifically for mixture distributions with a new sampling procedure. %Next, we transform the original samples into a latent space using quantile functions, which can also be difficult to evaluate, and perform conditioning in the latent space by using analytical formulas.

For applications, we will mainly focus on the Gaussian mixture copula model (GMCM) \cite{tewari2011parametric} due to its ease of implementation and its ability to capture multi-modal dependencies. Additionally, we demonstrate that incorporating automatic differentiation enhances the parameter fitting process of GMCM compared to conventional approaches for computing derivatives \cite{tewari2023estimation}.
We conduct numerical experiments using both simulated and real data to demonstrate the relevance of our approach, evaluating the results with (multivariate) scoring rules.

\begin{figure}[h]
    \centering
    \includegraphics[scale=0.85]{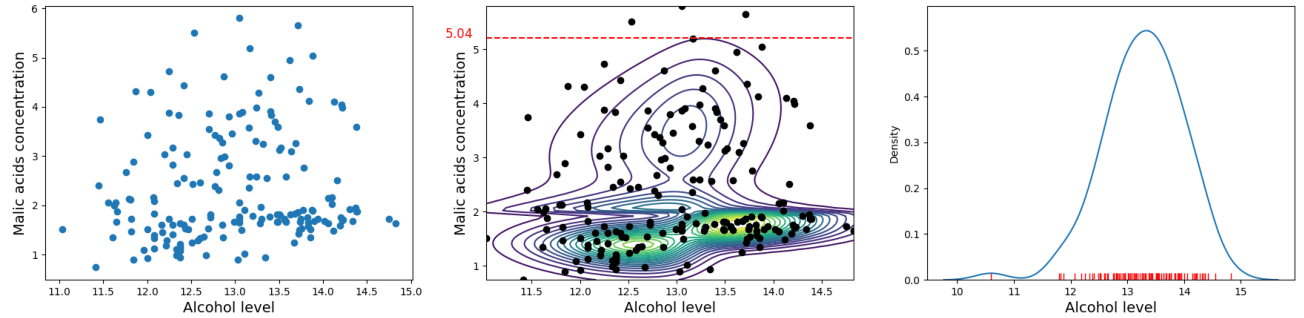}
    \caption{Steps of our conditional density estimation approach: from a sample of the joint distribution (left), we estimate the joint density by a mixture copula model  (middle) and for a given new value of Malic acids (5.04, red dotted line), we estimate / sample from the conditional distribution of Alcohol level (right) by leveraging easy conditioning in latent space.}
    \label{fig:gmcm_contours}
\end{figure}

As a first motivating example, we consider here the \textit{Wine data-set}, which represents chemical analysis of wines grown in the same region in Italy. The analysis determined the quantities of 13 constituents found in each of the three types of wines. For illustrative purposes, we focus on $2$ constituents, the Alcohol level and Malic acids concentration. Figure \ref{fig:gmcm_contours} illustrates the approach proposed using these variables. % which is to estimate the distribution of Alcohol level in a new wine given the value of its Malic acids concentration, from a joint sample previously acquired.
The joint density appears here to be rather well modeled by a GMCM, the conditional distribution of which can be estimated and simulated from %can be calculated and simulated from 
efficiently as detailed throughout the paper.

Our contributions can be summarized as follows:

\vspace{0.5em}

\begin{itemize}
    \item We introduce the concept of transdimensional families of distributions stable by conditioning. We provide various examples and counterexamples, demonstrating that this extends far beyond the Gaussian case, be it in the context of multivariate probability densities or of stochastic processes.

    \vspace{0.5em}
    
    \item We establish that this stability by conditioning is preserved for %finite
    mixtures and increasing transformations, supported by two key properties.

    \vspace{0.5em}
    
    \item Utilizing these insights, we develop a generative approach for sampling from conditional distributions by employing analytical formulas for conditioning. Building on the extension properties for %finite
    mixtures and increasing transformations, we apply this approach for the particular case of the Gaussian Mixture Copula Model.

    \vspace{0.5em}
    
    \item Through extensive numerical experiments on simulated and real-world data, we demonstrate that the use of non-Gaussian copulas allows for competitive scores in probabilistic prediction and imputation tasks while preserving desirable properties of a coherent probabilistic model.

    \vspace{0.5em}
    
\end{itemize}

We now review how the remainder of the paper is structured. 
%The remainder of the paper is organized as follows. 
In section \ref{sec:stable}, we introduce the properties of stability by marginalization and conditioning for families of multivariate distributions, provide some example and counter-examples families and show that these properties can be extended to mixtures and increasing transformations. In section \ref{sec::methods}, we %introduce the notion of "easy to condition" distribution, which is related to closure, and 
present our method for estimating efficiently a conditional distribution by assuming a parametric form for the joint distribution and leveraging analytical formulas for conditioning and present the particular case of the GMCM. %Section 4 focuses on the implementation of our algorithm for a particular case which is the GMCM. %, and present the evaluation metrics.
For the demonstration of the usefulness of our method, we perform numerical simulations and include real data examples in section \ref{sec:experiments}. Finally, in  section \ref{sec:conclusion}, we draw conclusions and outline areas for future research.

\section{Stability by conditioning}
\label{sec:stable}

\subsection{Multivariate distributions stable by conditioning}
\subsubsection{Definition}
%In this section, we will mainly work with probability density functions (pdf).

Let us first consider a fixed dimension $d\geq 1$ and a product of spaces $\prod_{k=1}^{d} \mathcal{X}_{k}$. The \(\mathcal{X}_k\) are assumed to be 
%\textcolor{red}{ordered} 
measure spaces, typically \(\mathbb{R}\) or intervals, each endowed with the Lebesgue measure or restrictions thereof.
We consider absolutely continuous probability distributions on $\prod_{k=1}^{d} \mathcal{X}_{k}$ with positive probability density functions $f^{(d)}(\cdot;\theta)$, where $\theta$ denotes a parameter. 
In this framework, we work indifferently with probability density functions (pdfs) or cumulative distribution functions (cdfs).

We denote as $\mathcal{F}^{(d)}= \{f^{(d)}(\cdot;\theta); \theta \in \Theta^{(d)}\}$ a parametric family of pdfs on $\prod_{k=1}^{d} \mathcal{X}_{k}$ with parameters belonging to a set $\Theta^{(d)}$. 
%Usually for applications we are interested in the case $\mathcal{X}_{k} = \mathbb{R}, ~\forall k \in \{1,\ldots,d\}$, which corresponds to considering probability distributions on $\mathbb{R}^{d}$.

We define a trans-dimensional family of pdfs by $\mathcal{F}= \bigcup \limits _{d=1}^{D} \mathcal{F}^{(d)}$ with $1\leq D  \leq +\infty$. The reason why we are concerned with this notion is because we will question the stability of such families by marginalization and conditioning with respect to some components.

Let us now introduce sets of indices that will be used for marginal distributions. For a given dimension $d>1$ %$d \geq 1$,
the sets of $\ell$ strictly ordered indices are: \begin{equation*}I_{\ell}^{(d)} = \{ \mathbf{i}= (i_{1},\ldots,i_{\ell}), ~ 1 \leq i_{1} < \ldots < i_{\ell} \leq d \}, ~\text{with}~ 1\leq\ \ell \leq d.
\end{equation*}

For conditional distributions, we define a set of pairs of strictly ordered indices $(\mathbf{i}, \mathbf{j}) \in  I_{\ell,m}^{d}$ with elements $\mathbf{i}$ and $\mathbf{j}$ satisfying:
\begin{itemize}
    \item $\mathbf{i} = (i_{1},\ldots,i_{\ell})$ set of $\ell$ indices with $1 \leq i_{1}  < \ldots < i_{\ell} \leq d$, $1\leq \ell \leq d-1$, 
    \item $\mathbf{j}=(j_{1},\ldots, j_{m})$ set of $m$ indices with $1\leq j_{1} < \ldots < j_{m} \leq d$, $1 \leq m \leq d-\ell$,
    \item  $\text{set}(\mathbf{i}) \cap \text{set}(\mathbf{j}) = \emptyset $ where the set of a vector is the set of all its components.
\end{itemize}

For $f$ a positive pdf on $\prod_{k=1}^{d}\mathcal{X}_{k}$, $1\leq \ell\leq d-1$, and %a set of $\ell$ strictly ordered indices $\mathbf{i} \in I_{\ell}^{(d)}$, 
$(\mathbf{i}, \mathbf{j}) \in  I_{\ell,d-\ell}^{d}$,
we define the function:
$$f_{\mathbf{i}} : (x_{i_{1}},\ldots,x_{i_{\ell}})\in \prod_{k=1}^{\ell} \mathcal{X}_{i_{k}} \rightarrow \int_{\prod_{k=1}^{d-\ell} \mathcal{X}_{j_{k}}} f(x) dx_{j_{1}} \ldots dx_{j_{d-\ell}},$$
which applies the marginalization of a $d$-dimensional density over $d-\ell$ variables. Let us note that since $\mathbf{j}$ has $d-\ell$ components, it is completely defined by $\mathbf{i}$. 

\vspace{1em}

To extend this operation to a trans-dimensional setting, we extend the marginalization operator as follows. 
%This operator takes as input an element of a parametric family of pdfs and returns all possible marginal distributions associated with it.
For $d\geq 1$ and $f^{(d)}(\cdot;\theta) \in \mathcal{F}^{(d)}$, we define :
$$\text{Marg}(f^{(d)}(\cdot;\theta)) = \bigcup\limits_{\ell=1}^{d-1} \{f_{\mathbf{i}}(\cdot;\theta), ~\mathbf{i} \in I_{\ell}^{(d)}\}. $$
Now if we apply $\text{Marg}$ to the entire family of distributions $\mathcal{F}^{(d)}$, we get:
$$\text{Marg}(\mathcal{F}^{(d)}) = \left \{ \text{Marg}(f^{(d)}(\cdot;\theta)), \theta \in \Theta^{(d)}\right \},$$ and finally for a trans-dimensional family of distributions $\mathcal{F}$ we set:
$$\text{Marg}(\mathcal{F}) = \bigcup\limits_{d=1}^{D} \text{Marg}(\mathcal{F}^{(d)}).$$

We say that the trans-dimensional family of probability distributions $\mathcal{F}$ is stable by marginalization if it satisfies $\text{Marg}(\mathcal{F}) \subset \mathcal{F}$. Although the main focus of the paper is on conditioning, all subsequent results are also applicable to marginalization.

\vspace{1em}

Let us now consider conditioning. Since the pdfs considered here are positive, all considered conditional distributions are well-defined.
For $f$ a positive pdf on $\prod_{k=1}^{d}\mathcal{X}_{k}$, a pair of sets of strictly ordered indices $(\mathbf{i},\mathbf{j}) \in  I_{\ell,m}^{d}$ and observed components $(x_{j_{1}},\ldots,x_{j_{m}})$, we denote:

$$f_{\mathbf{i}|\mathbf{j}} (\cdot|(x_{j_{1}},\ldots,x_{j_{m}})) : ( x_{i_{1}},\ldots, x_{i_{\ell}})  \in \prod_{k=1}^{\ell} \mathcal{X}_{i_{k}}  \rightarrow \frac{f_{\left [\mathbf{i},\mathbf{j}\right]} (x_{i_{1}}, \ldots,x_{i_{\ell}}, x_{j_{1}},\ldots,x_{j_{m}})}{f_{\mathbf{j}}(x_{j_{1}},\ldots,x_{j_{m}})},$$
%$$f_{\mathbf{i}|\mathbf{j}} ((x_{i_{1}},\ldots,x_{i_{l}})|(x_{j_{1}},\ldots,x_{j_{m}})) = \frac{f_{\left [\mathbf{i},\mathbf{j}\right]} (x_{i_{1}}, \ldots,x_{i_{l}}, x_{j_{1}},\ldots,x_{j_{m}})}{f_{\mathbf{j}(x_{j_{1}},\ldots,x_{j_{m}})}}$$

where $\left [\mathbf{i},\mathbf{j}\right ] = (i_{1},\ldots,i_{\ell},j_{1},\ldots,j_{m})$.

In the same manner as above for marginalization, we define a conditioning operator taking a pdf in a parametric family as input and returning all possible conditional densities i.e for all pairs of sets of strictly ordered indices and every possible set of conditioning variables. For a function $f^{(d)}(\cdot;\theta) \in \mathcal{F}^{(d)}$, we have : $$\text{Cond}(f^{(d)}(\cdot,\theta)) = \left\{ f^{(d)}_{\mathbf{i}|\mathbf{j}}(\cdot|(x_{j_{1}}, \ldots,x_{j_{m}});\theta),~(\mathbf{i},\mathbf{j}) \in I_{\ell,m}^{d} ,~ (x_{j_{1}},\ldots, x_{j_{m}}) \in \prod_{k=1}^{m} \mathcal{X}_{j_{k}}\right \}.$$

By applying it to the trans-dimensional family of pdf $\mathcal{F}$ we get:
$$\text{Cond}(\mathcal{F}) =  \bigcup \limits_{d=1}^{D}\left\{\text{Cond}(f^{(d)}(\cdot;\theta)),~\theta \in \Theta^{(d)}\right\}$$

We say that trans-dimensional family of probability distributions $\mathcal{F}$ is stable by conditioning if it satisfies  $\text{Cond}(\mathcal{F}) \subset \mathcal{F}$.

\begin{remark}
    While Eaton \cite{eaton1989group} studied the invariance of families of probability distributions under group actions, the corresponding notion of stability conditioning of multivariate distributions has, to the best of our knowledge, not been explicitly defined in the literature. Similar to conditionally specified distributions \cite{arnold2012conditionally}, this stability property would enable the specification of all conditional distributions. 
\end{remark}

\subsubsection{Examples}

Stability by marginalization and conditioning can be demonstrated by showing that the marginal or conditional distributions remain within the same family of distributions, albeit in a lower dimension. Here, we will present several examples of distribution families that exhibit this property, as well as two counter-examples.

\vspace{0.5em}

\begin{itemize}

 \item \textbf{Gaussian distribution}: For a fixed dimension $d \geq 1 $, we consider 
 %the family of Gaussian distributions on $\mathbb{R}^{d}$ by: 
 $$\mathcal{F}^{(d)} = \left\{ f^{(d)} (x_{1},\ldots,x_{d}; \mu, \Sigma) = \frac{\exp(-\frac{1}{2}(\mathbf{x}-\mu)^{T} \Sigma^{-1}(\mathbf{x}-\mu)) }{\sqrt{(2\pi)^{d} |\Sigma|}}, \theta = (\mu, \Sigma) \in \Theta^{(d)}\right\},$$ where $\Theta^{(d)} = \mathbb{R}^{d} \times \text{PD}_{d}$ where $\text{PD}_{d}$ is the set of all symmetric positive definite matrices on $\mathbb{R}^{d \times d} $.
The trans-dimensional family of Gaussian distributions is defined by: $\mathcal{F} = \bigcup \limits_{d=1}^{D} \mathcal{F}^{(d)}$ with $D= + \infty$. 
By properties of the multivariate Gaussian distribution, we know that associated marginal and conditional distributions remain Gaussian with parameters that can be computed analytically via well-known formulas. %we know that if the joint distribution $f_{\left [\mathbf{i},\mathbf{j}\right]}$ is multivariate Gaussian with parameters $\mu$ and $\Sigma$ then its marginal distributions $f_{\mathbf{i}} $ is Gaussian with parameters $\mu_{\mathbf{i}}$ and $\Sigma_{\mathbf{i} \mathbf{i}}$. 
We have that $\text{Marg}(\mathcal{F}) \subset \mathcal{F}$ and  $\text{Cond}(\mathcal{F}) \subset \mathcal{F}$, and thus the trans-dimensional family of Gaussian distributions is stable by marginalization and by conditioning.

%Furthermore, we also know that $f_{\mathbf{i}|\mathbf{j}}$ is multivariate Gaussian with parameters $\mu_{\mathbf{i}|\mathbf{j}} = \mu_{\mathbf{i}} + \Sigma_{\mathbf{i}\mathbf{j}} \Sigma_{\mathbf{j}\mathbf{j}}^{-1}(\textbf{x}_{\mathbf{j}}-\mu_{\mathbf{j}})$ and $\Sigma_{\mathbf{i}|\mathbf{j}} = \Sigma_{\mathbf{i}\mathbf{i}}-\Sigma_{\mathbf{i}\mathbf{j}} \Sigma_{\mathbf{j}\mathbf{j}}^{-1} \Sigma_{\mathbf{j}\mathbf{i}}$. 
%Then we have that $\text{Cond}(\mathcal{F}) \subset \mathcal{F}$, the trans-dimensional family of Gaussian distributions is stable by conditioning.

\vspace{1em}

\item  \textbf{Student $t$-distribution}: Another example is the family of multivariate Student t-distributions. %which generalizes multivariate normal distributions (appearing as limiting case) while allowing for heavier tails. 
Unlike the multivariate Gaussian distribution, it can model tail dependencies and it includes an additional “degrees of freedom” parameter, which controls the shape of the distribution, with higher values leading to distributions that resemble the multivariate Gaussian, while lower values result in heavier tails.

It is defined as a scaled mixture of Gaussian distributions. In particular, if $\textbf{Y} \sim \mathcal{N}_{d}(0,\Sigma)$ and $u\sim \chi^{2}_{\nu}$ are independent then $ \textbf{X} = \frac{\textbf{Y}}{\sqrt{\frac{u}{\nu}}} + \mu$ is such that $\textbf{X} \sim t_{d}(\mu,\Sigma,\nu)$, where $\chi^{2}_{\nu}$ represents a chi-square distribution with $\nu$ degrees of freedom. We say that $\textbf{X}$ follows a multivariate $t$-distribution with mean $\mu$, covariance matrix $\Sigma$ and $\nu$ degrees of freedom.

For $d \geq 1$, we denote the family of  multivariate Student $t$-distribution as:
\begin{flalign*}
\mathcal{F}^{(d)} = \left\{ f^{(d)} (\mathbf{x}; \mu, \Sigma, \nu), \theta = (\mu, \Sigma, \nu) \in \Theta^{(d)}=\mathbb{R}^{d} \times \text{PD}_{d} \times  \left[ 0, +\infty \right[ \right\},
\end{flalign*}
with $f^{(d)} (\mathbf{x}; \mu, \Sigma, \nu) = \frac{\Gamma(\frac{\nu +d}{2})}{\Gamma(\frac{\nu}{2}) (\nu \pi)^{\frac{d}{2}} |\Sigma|^{\frac{1}{2}}}(1 + \nu ^{-1}(\mathbf{x}-\mu)^{T} \Sigma^{-1} (\mathbf{x}-\mu))^{-\frac{(\nu +d)}{2}}$.
%and $$.

The multivariate Student $t$-distribution is stable by marginalization and conditioning \cite{ding2016conditional}; more details are included in the appendix.

\vspace{1em}

\item \textbf{Elliptical distributions: } Elliptical distributions are characterized by elliptical contours and provide a flexible framework for modeling multivariate data with different scales, orientations, and dependence structures across dimensions. They encompass the two earlier examples of Gaussian and Student $t$-distribution. A $d$-dimensional random vector $\mathbf{X}$ has an elliptical distribution with location $\mu$ and dispersion $\Sigma$ if it can be written: $\mathbf{X}=\mu + R_{d}\mathbf{A} \mathbf{U}_{d}$ with $\mathbf{U}_d$ uniformly distributed on the $d$ dimensional unit sphere $\mathbb{S}_{d}$, $\mathbf{A}\in\mathbb{R}^{d\times d}$ is a matrix such that
$\mathbf{A}\mathbf{A}^{\top}=\Sigma$ and $R_{d}$ an independent scalar random variable. If the density of $R_{d}$ in the stochastic representation exists, and $\Sigma$ positive definite, then $h$ the density of $\mathbf{X}$ can be expressed by: \[ h(\mathbf{x};\mu,\Sigma,g_{d}) =|\Sigma| ^{-\frac{1}{2}} g_{d}((\mathbf{x}-\mu)^{t}  \Sigma^{-1} (\mathbf{x}-\mu))
    \]
    with $g_{d}$ uniquely determined by the distribution of $R_{d}$ and called the density generator. We note that $X$ has distribution $\mathcal{E}_{d}(\mu,\Sigma, g_{d})$.
    The class of elliptical distributions is stable under marginalization; however, the marginal distribution does not necessarily belong to the same parametric elliptical family as the original distribution \cite{wang2013practical}. Kano established necessary and sufficient conditions on the density generator for the marginal consistency of an elliptical family \cite{kano1994consistency}. The conditional distribution of an $l$-dimensional vector $\mathbf{X_{1}}$ given $\mathbf{X_{2}}=\mathbf{x_{2}}$ is still an elliptical distribution but potentially with another density generator than $X$ \cite{fang2018symmetric}. Indeed, we have for \(\mathbf{\mu}= 0\) that the conditional distribution is \(\mathcal{E}_l(\mu_{1.2}, \Sigma_{11.2}, g_{l,d,a}(\mathbf{x}_2))\), where \(\mu_{1.2} = \Sigma_{12}\Sigma_{22}^{-1}\mathbf{x}_2\), \(\Sigma_{11.2} = \Sigma_{11} - \Sigma_{12}\Sigma_{22}^{-1}\Sigma_{21}\), and the density generator \(g_{l,d,a}\) is

$$
g_{l,d,a}(u) = \frac{\Gamma(r/2) g_{d}(a+u)}{\pi^{r/2} \int_{0}^{\infty} v^{r/2-1} g_{d}(a+v)  dv}.
$$
with $a = \mathbf{x}_{2}^{t} \mathbf{x}_{2}$. Conditioning an elliptical distribution can alter its generator,
and a closed-form expression for the conditional distribution is generally not available. This limitation restricts its applicability in the remainder of this paper. 

\vspace{1em}

\item \textbf{Other examples}: Other examples of families of multivariate distributions stable by marginalization and conditioning are the multivariate unified skew normal and skew Student $t$-distributions \cite{azzalini2013skew}.
%The family of elliptically contoured distributions also enjoys these properties \cite{cambanis1981theory}.  However, as shown in \cite{fang2018symmetric}, conditioning an elliptical distribution can alter its generator, and a closed-form expression for the conditional distribution is generally not available. This limitation restricts its applicability in the remainder of this paper. 
On a different note, Dirichlet distributions are stable by conditioning, up to scalings ensuring that resulting measures are supported on the unit simplex (in appropriate dimensions) \cite{fang2018symmetric}. Furthermore, Liouville distributions \cite{mcneil2010archimedean}, which are mixtures of Dirichlet distributions, exhibit stability by conditioning, as demonstrated in Fang's book \cite{fang2018symmetric}, Section 6.4. Additionally, it is worth mentioning that Archimedean copulas are also stable by conditioning and marginalization, as discussed in Joe's book \cite{joe2014dependence}, Section 3.2.  %Unlike elliptical distributions, they allow for skewness.

\vspace{1em}

\item \textbf{Counter example 1}: Student $t$-distribution with fixed degrees of freedom parameter.

Let us now consider a simple counter-example of a family of multivariate distributions which is stable by marginalization but not stable by conditioning.
For a given $d \geq 1$, consider the 
family of multivariate Student $t$-distributions with a fixed degree of freedom $\nu \in  \left[ 0, +\infty \right[$ i.e 
\begin{flalign*}
\mathcal{F}^{(d)}_{\nu} =  \left\{ f^{(d)} (\mathbf{x}; \mu, \Sigma),\theta = (\mu, \Sigma) \in \Theta^{(d)}_{\nu}=\mathbb{R}^{d} \times \text{PD}_{d} \right\},
\end{flalign*}
where $f^{(d)} (\mathbf{x}; \mu, \Sigma)= \frac{\Gamma(\frac{\nu +d}{2})}{\Gamma(\frac{\nu}{2}) (\nu \pi)^{\frac{d}{2}} |\Sigma|^{\frac{1}{2}}}\left (1 + \nu ^{-1}(\mathbf{x}-\mu)^{T} \Sigma^{-1} (\mathbf{x}-\mu)\right)^{-\frac{(\nu +d)}{2}}.$ 

The transdimensional family is defined by $\mathcal{F}_{\nu} = \bigcup \limits_{d=1}^{D} \mathcal{F}^{(d)}_{\nu}$ with $D = + \infty$.
Then, by properties of the multivariate Student $t$-distribution \cite{ding2016conditional}, this family is stable by marginalization. However, it is not stable by conditioning, because the resulting Student conditional distributions possess degrees of freedom different from $\nu$.

\vspace{0.5em}

\item \textbf{Counter example 2}: $q$-Exponential distributions.

A multivariate \( q \)-exponential distribution \cite{li2023bayesian}, denoted as \( q\text{-ED}_{d}(\boldsymbol{\mu}, \mathbf{C}) \) for  $(\boldsymbol{\mu}, \mathbf{C}) \in \mathbb{R}^{d} \times \text{PD}_{d}$, has the following density for $\mathbf{x} \in \mathbb{R}^{d}$ ($d\geq 1$ being fixed, here):
\[
p(\mathbf{x} \mid \boldsymbol{\mu}, \mathbf{C}, q) = \frac{q}{2} (2\pi)^{-\frac{d}{2}} |\mathbf{C}|^{-\frac{1}{2}} r^{\frac{q}{2}-1}\exp \left( -\frac{r^{\frac{q}{2}}}{2} \right ), \quad r(\mathbf{x}) = (\mathbf{x} - \boldsymbol{\mu})^{\top} \mathbf{C}^{-1} (\mathbf{x} - \boldsymbol{\mu}).
\]
This family of distributions is stable by marginalization by using Kano's \cite{kano1994consistency} conditions, however for a given index $q$, the corresponding family of distributions is not stable by conditioning. 
From \cite{li2023bayesian}, we know that the $q$-exponential distribution is consistent in the sense defined by Kano \cite{kano1994consistency}. By considering, $d=2$, $\mu = (0,0)^{t}$, $C=I_{2}$ and $\mathbf{x} = (x_{1},x_{2})^{t}$, we then have for $x_{1} \in \mathbb{R}$ :
$$p(x_{1}) \propto (x_{1}^{2})^{{\frac{q}{2}}-1} \exp{\left(-\frac{1}{2} x_{1}^{q}\right)}.$$
 By assuming $p(x_{1}) \neq 0$, the conditional distribution can then be written: 
$$p(x_{2}|x_{1}) \propto \frac{p(x_{1},x_{2})}{p(x_{1})} \propto \frac{(x_{1}^{2}+x_{2}^{2})^{(\frac{q}{2}-1)}}{(x_{1}^{2})^{(\frac{q}{2}-1)}} \exp{\left(-\frac{(x_{1}^{2}+x_{2}^{2})^{\frac{q}{2}}}{2} + \frac{(x_{1}^{2})^{\frac{q}{2}}}{2}\right)}.$$

By taking $q=4$ and focusing on the exponential term, we get that the exponential term is:
$$\exp{\left(-\frac{x_{2}^{4}-2x_{1}^{2}x_{2}^{2}}{2}\right)}.$$

If the conditional distribution were $q$-exponential (with $q=4$) it would be possible to write the exponential term as $\exp(-\frac{r^{2}}{2})$ with $r$ a quadratic form in $x_{2}$, which is not the case. 
%However the term derived above cannot be written as a square of a quadratic term in $r$. 
We conclude that the conditional distribution is not $q$-exponential with $q=4$.
\end{itemize}

%\textbf{Remark}: Up to a scaling factor, the Dirichlet distribution is stable by conditioning \cite{fang2018symmetric}. %In general, exponential families of distributions aren't stable by conditioning, however when the base reference measure depends only on the conditioned variables.
\subsection{Main properties}
\subsubsection{Extension to finite mixtures}

%Let us first define a “mixing” operator which takes as input a parameterized family of multivariate pdfs with a fixed dimension $d\geq 1$ and returns the set of probabilities densities that are finite mixtures of elements of this family.

%\vspace{0.5em}

Consider again a parametrized family of pdfs %$\mathcal{F^{(d)}}$ 
denoted $\mathcal{F}^{(d)}= \{f^{(d)}(\cdot;\theta); \Theta \in \Theta^{(d)}\}$. We define 
%the mixing operator as:

\begin{flalign*}
\small{
\text{Mix}(\mathcal{F}^{(d)}) = \left \{ 
%f(\cdot)= 
\sum_{k=1}^{K} \alpha_{k} f^{(d)}(\cdot;\theta_{k}),~ K\geq 1, ~ \theta_{1},\ldots,\theta_{K} \in \Theta^{(d)} , ~\alpha_{1},\ldots,\alpha_{K} \geq 0 ~ \text{s.t} ~ \sum_{k=1}^{K} \alpha_{k} =1  \right\}}.
 \end{flalign*}

For a trans-dimensional family of distributions $\mathcal{F}$, we define its mixing set of distributions as the union over the mixing sets of the families with fixed dimensions i.e
$$\text{Mix}(\mathcal{F}) = \bigcup \limits_{d=1}^ {D} \text{Mix}(\mathcal{F}^{(d)}).$$
We now present results that extend the concepts of stability by marginalization and conditioning to finite mixtures.

\vspace{1em}

\begin{theorem}
\label{thm:extension_mixture}
    If a parametrized family of trans-dimensional probability distributions $\mathcal{F}$ is stable by conditioning (resp. marginalization) then  $\text{Mix}(\mathcal{F})$ is also stable by conditioning (resp. marginalization).
\end{theorem}

\begin{proof}[Proof of Theorem \ref{thm:extension_mixture}] Let $f$ in $\text{Mix}(\mathcal{F})$. There exists $d \in \{1,\ldots,D\}$, $K\geq 1$, $\theta_{1},\ldots,\theta_{K} \in \Theta^{(d)}$, $\alpha_{1},\ldots,\alpha_{K} \geq 0 ~ \text{s.t} ~ \sum_{k=1}^{K} \alpha_{k} =1$ such that:
$$f(\cdot) = \sum_{k=1}^{K} \alpha_{k} f^{(d)}(\cdot;\theta_{k}).$$
%such that $f \in \text{Cond}(f^{(d)}(\cdot,\psi))$ with $f^{(d)}(\cdot,\psi) \in \text{Mix}(\mathcal{F}^{(d)})$. We can write $f^{(d)}(\cdot,\psi)) = \sum_{k=1}^{K} \alpha_{k} f^{(d)}(\cdot,\psi_{k}) $ for $K \geq 1$, $\psi_{1},\ldots,\psi_{K} \in \Psi^{(d)}$ and $\alpha_{1},\ldots,\alpha_{K} \geq 0 ~ \text{s.t} ~ \sum_{k=1}^{K} \alpha_{k} =1$. 
For every pair of strictly ordered indices $(\mathbf{i}, \mathbf{j}) \in  I_{\ell,m}^{d}$ and conditioning variables $(x_{j_{1}},\ldots,x_{j_{m}}) \in \prod_{k=1}^{m} \mathcal{X}_{j_{k}}$, we have that $\forall (x_{i_{1}},\ldots,x_{i_{\ell}}) \in \prod_{k=1}^{\ell} \mathcal{X}_{j_{k}}$:
\begin{equation*}
\begin{split}
 f_{\mathbf{i}|\mathbf{j}} (( x_{i_{1}},\ldots, x_{i_{\ell}})|(x_{j_{1}},\ldots,x_{j_{m}})) &= \frac{f_{[\mathbf{i},\mathbf{j}]}( x_{i_{1}},\ldots, x_{i_{\ell}},x_{j_{1}},\ldots,x_{j_{m}})}{f_{\mathbf{j}}(x_{j_{1}},\ldots,x_{j_{m}})} \\
 &= \frac{1}{f_{\mathbf{j}}(x_{j_{1}},\ldots,x_{j_{m}})} \sum_{k=1}^{K} \alpha_{k} f^{(d)}_{[\mathbf{i},\mathbf{j}]}(x_{i_{1}},\ldots, x_{i_{\ell}},x_{j_{1}},\ldots,x_{j_{m}}; \theta_{k}) \\
 &= \sum_{k=1}^{K} \frac{\alpha_{k} f^{(d)}_{\mathbf{j}}(x_{j_{1}},\ldots,x_{j_{m}};\theta_{k})}{f_{\mathbf{j}}(x_{j_{1}},\ldots,x_{j_{m}})} f^{(d)}_{\mathbf{i}|\mathbf{j}}(( x_{i_{1}},\ldots, x_{i_{\ell}})|(x_{j_{1}},\ldots,x_{j_{m}}); \theta_{k}) \\
 &= \sum_{k=1}^{K} \Tilde{\alpha}_{k}(x_{j_{1}},\ldots,x_{j_{m}}) f^{(d)}_{\mathbf{i}|\mathbf{j}}(( x_{i_{1}},\ldots, x_{i_{\ell}})|(x_{j_{1}},\ldots,x_{j_{m}}); \theta_{k}),
\end{split}
\end{equation*}
%\begin{equation*}
%\begin{split}
 %f_{\mathbf{i}|\mathbf{j}} ( ( x_{i_{1}},\ldots, x_{i_{\ell}})|(x_{j_{1}},\ldots,x_{j_{m}})) &= \frac{f_{[\mathbf{i},\mathbf{j}]}( x_{i_{1}},\ldots, x_{i_{\ell}},x_{j_{1}},\ldots,x_{j_{m}})}{f_{\mathbf{j}}(x_{j_{1}},\ldots,x_{j_{m}})}\\ &= \frac{1}{f_{\mathbf{j}}(x_{j_{1}},\ldots,x_{j_{m}})} \sum_{k=1}^{K} \alpha_{k} f^{(d)}_{[\mathbf{i},\mathbf{j}]}(x_{i_{1}},\ldots, x_{i_{\ell}},x_{j_{1}},\ldots,x_{j_{m}}; \theta_{k})% \\ &= \frac{1}{f^{(d)}_{\mathbf{j}}(x_{j_{1}},\ldots,x_{j_{m}};\psi)}  \sum_{k=1}^{K} \alpha_{k} f^{(d)}_{\mathbf{j}}(x_{j_{1}},\ldots,x_{j_{m}};\psi_{k}) f^{(d)}_{\mathbf{i}|\mathbf{j}}(( x_{i_{1}},\ldots, x_{i_{\ell}})|(x_{j_{1}},\ldots,x_{j_{m}});\psi_{k}) 
 %\\&=  \sum_{k=1}^{K} \frac{\alpha_{k} f^{(d)}_{\mathbf{j}}(x_{j_{1}},\ldots,x_{j_{m}};\theta_{k})}{f_{\mathbf{j}}(x_{j_{1}},\ldots,x_{j_{m}})}  f^{(d)}_{\mathbf{i}|\mathbf{j}}(( x_{i_{1}},\ldots, x_{i_{\ell}})|(x_{j_{1}},\ldots,x_{j_{m}}); \theta_{k}) 
    %\\&= \sum_{k=1}^{K} \Tilde{\alpha}_{k}(x_{j_{1}},\ldots,x_{j_{m}}) f^{(d)}_{\mathbf{i}|\mathbf{j}}(( x_{i_{1}},\ldots, x_{i_{\ell}})|(x_{j_{1}},\ldots,x_{j_{m}}); \theta_{k})
%\end{split}
%\end{equation*}
where $\Tilde{\alpha_{k}}(x_{j_{1}},\ldots,x_{j_{m}})=\frac{\alpha_k f^{(d)}_{\mathbf{j}}(x_{j_{1}},\ldots,x_{j_{m}};\theta_{k})}{f_{\mathbf{j}}(x_{j_{1}},\ldots,x_{j_{m}})}$ which depends only on indices in $\mathbf{j}$.

Since $\text{Cond}(\mathcal{F})\subset \mathcal{F}$, then for every $k \in \{1,\ldots,K\}, ~f^{(d)}_{\mathbf{i}|\mathbf{j}}(\cdot|(x_{j_{1}},\ldots,x_{j_{m}}); \theta_{k}) \in \mathcal{F}$ and the conditional distribution $f_{\mathbf{i}|\mathbf{j}}$ is a mixture of elements of $\mathcal{F}$ with weights $\Tilde{\alpha}_{k}$ so we can conclude that $f_{\mathbf{i}|\mathbf{j}} \in \text{Mix}(\mathcal{F})$. The proof for marginalization is analogous. 
\end{proof}%\hfill $\square$

\begin{remark}
    This result extends to infinite mixtures, with a similar proof provided in Appendix \ref{sec:infinite}. However, when dealing with infinite mixtures, computations become intractable, and analytical conditioning formulas, crucial for our approach in Section \ref{sec::methods}, are unavailable. Consequently, the practical utility of this extension to infinite mixtures remains limited.
\end{remark}

\vspace{1em}
This result can similarly be applied to cdfs. %(cdf). In the rest, we will work indifferently with pdf or cdf to represent continuous distributions.
Note that if we can compute analytically the conditioned mixing weights $\Tilde{\alpha}_{k}$ and if we can sample from the conditional distributions of each component $f_{\mathbf{i}|\mathbf{j}}(\cdot;\theta_{k})$, we can produce samples from the conditionals of the overarching mixture of distributions. Although the authors are not aware of prior formulations of Theorem~\ref{thm:extension_mixture}, it its worth remarking that the formula for the conditional distribution of a mixture of distributions has in fact been utilized in diverse contexts \cite{gilardi2002conditional,spycher2009multivariate}.

\subsubsection{Extension to increasing transformations}

We now work with parametric families of cdfs on $\prod_{k=1}^{d} \mathcal{X}_{k}$, similarly denoted $\mathcal{F}^{d} = \left \{ F^{(d)}(\cdot;\theta); \theta \in \Theta^{(d)}\right\}= \left \{ F^{(d)}_{\theta}; \theta \in \Theta^{(d)}\right\}$ as done with pdfs in the previous section. Similarly, we denote resulting trans-dimensional families as  $\mathcal{F} = \bigcup \limits_{d=1}^{D} \mathcal{F}^{(d)}$ with $D \leq +\infty$.
%
%Let define a "push-forward" operator which takes as input an element of a parametric family of cumulative distribution functions and returns every possible cumulative distribution functions obtained by applying marginal $\mathcal{C}^{1}$-diffeomorphisms to the components. 
%
For fixed  $d\geq 1$ and $\mathcal{X}'_{i} (i=1,\ldots,d$) %\textcolor{red}{ordered} 
measure spaces,  we further define 
%the class of functions: 
$$\Xi^{d} = \left \{  \xi = \begin{pmatrix}
    \xi_{1} \\ \vdots \\ \xi_{d}
\end{pmatrix} ~ \text{s.t}~ \xi_{i} \in \mathcal{C}^{1}_{\uparrow}(\mathcal{X}_{i},\mathcal{X}'_{i}) ,~i =1,\ldots,d \right \},$$ where $\mathcal{C}^{1}_{\uparrow}(\mathcal{X}_{i},\mathcal{X}'_{i})$ is the class of increasing $\mathcal{C}^{1}$ diffeomorphisms from $\mathcal{X}_{i}$ to $\mathcal{X}'_{i}$. We introduce a “pushforward” operator, which takes as input an element of a parametric family of cdfs and returns all possible cdfs obtained by applying marginal $\mathcal{C}^{1}_{\uparrow}$ diffeomorphisms to the components, by: 

$$\Xi^{d} \# F^{(d)}_{\theta} =\left \{ F^{(d)}_{\theta} \circ \xi^{-1}; \xi \in \Xi^{d}\right \}.$$

Note that elements of $\Xi^{(d)} \# F^{(d)}_{\theta}$ are probability distributions on $\prod_{i=1}^{d} \mathcal{X'}_{i}$ and that %we restricted to differentiable bijections such that
the obtained probability distributions are still absolutely continuous.
%$$\text{Meta}(F^{(d)}(\cdot;\psi)) = \left \{ F(\cdot) = F^{(d)} (\phi_{\mathbf{i}}(x_{i_{1}},\ldots,x_{i_{\ell}}),\phi_{\mathbf{j}}(x_{j_{1}},\ldots,x_{j_{m}});\psi); (\mathbf{i},\mathbf{j}) \in I_{\ell,m}^{(d)} ~\text{with}~ \ell + m =d, ~\phi_{\mathbf{i}}, \phi_{\mathbf{j}}~ \text{bijections}\right \}$$

%$$\text{Meta}(F^{(d)}(\cdot;\psi)) = \left \{ F(\cdot) = (F^{(d)} \circ \phi) (\cdot; \psi); ~\phi = \begin{pmatrix}
   % \phi_{1} \\ \ldots \\ \phi_{d}
%\end{pmatrix} \in \Phi^{d}  \right \}$$
%where $\Phi^{d} = \left \{  \phi = \begin{pmatrix}
    %\phi_{1} \\ \ldots \\ \phi_{d}
%\end{pmatrix} ~ \text{s.t}~ \phi_{i} \in \mathcal{B}(\mathcal{X}_{i}) ,~i =1,\ldots,d \right \} $
%with $\mathcal{B}(\mathcal{X}_{i})$ is the set of bijections from $\mathcal{X}_{i}$ to itself.
%where $\phi_{\mathbf{i}}$ is a bijection from $\prod_{k=1}^{\ell} \mathcal{X}_{i_{k}}$ to itself and $\phi_{\mathbf{j}}$ is a bijection from $\prod_{k=1}^{m} \mathcal{X}_{j_{k}}$ to itself.
By applying the operator to the whole family of cdfs $\mathcal{F}^{(d)}$ we get:

$$\Xi^{d} \# \mathcal{F}^{(d)} = \left \{ \Xi^{d} \# F^{(d)}_{\theta}; \theta \in \Theta^{(d)} \right\}.$$

Defining finally the class of functions with increasing marginal $\mathcal{C}^{1}$ diffeomorphisms as $$\Xi = \bigcup \limits_{d=1}^{D} \Xi^{d},$$ we can now apply the pushforward operator to a trans-dimensional family of cdfs $\mathcal{F}$ by:
$$\Xi \# \mathcal{F} = \bigcup \limits_{d=1}^{D} \Xi^{d} \# \mathcal{F}^{(d)}.$$

We call the trans-dimensional family $\Xi \#\mathcal{F}$ the pushforward of $\mathcal{F}$. In the next section, we relate this concept to copulas and meta families of distributions. At this point, we extend the stability of marginalization and conditioning results to transformations.
%For example, when $\mathcal{F}$ is the trans-dimensional family of Gaussian distributions, we say that $\text{Meta}(\mathcal{F})$ is the meta family of Gaussian distributions.

\begin{theorem}
\label{thm::transfo}
    If a parametrized family of trans-dimensional probability distributions $\mathcal{F}$ is stable by conditioning (resp. marginalization) then 
    %its push-forward family 
    $\Xi \#\mathcal{F}$ is also stable by conditioning (resp. marginalization).
\end{theorem}

\begin{proof}[Proof of Theorem \ref{thm::transfo}]: Let $F \in \Xi \# \mathcal{F}$. There exists $d \in \{1,\ldots,D\},~\theta \in \Theta^{(d)}, ~\xi \in \Xi^{d}$ such that $\forall (x_{1},\ldots,x_{d}) \in \prod_{k=1}^{d} \mathcal{X}_{k}$:
$$ F(x_{1},\ldots,x_{d}) = (F^{(d)} \circ \xi^{-1})(x_{1}, \ldots,x_{d}; \theta) = F^{(d)}(\xi_{1}^{-1}(x_{1}),\ldots,\xi_{d}^{-1}(x_{d}); \theta).$$

For every pair of strictly of ordered indices $(\mathbf{i}, \mathbf{j}) \in  I_{\ell,m}^{d}$ and conditioning variables 

$(x_{j_{1}},\ldots,x_{j_{m}}) \in \prod_{k=1}^{m} \mathcal{X}_{j_{k}}$, we have that $\forall (x_{i_{1}},\ldots,x_{i_{\ell}}) \in \prod_{k=1}^{\ell} \mathcal{X}_{j_{k}}$:

\begin{equation*}
    \begin{split}
         F_{\mathbf{i}|\mathbf{j}} (x_{i_{1}},\ldots, x_{i_{\ell}}|x_{j_{1}},\ldots,x_{j_{m}}) &=  F^{(d)}_{\mathbf{i}|\mathbf{j}}(\xi_{i_{1}}^{-1}(x_{i_{1}}),\ldots,\xi_{i_{\ell}}^{-1}(x_{i_{\ell}})|\xi_{j_{1}}^{-1}(x_{j_{1}}),\ldots,\xi_{j_{m}}^{-1}(x_{j_{m}}) ; \theta) \\ &=
         F^{(d)}_{\mathbf{i}|\mathbf{j}}(z_{i_{1}},\ldots, z_{i_{\ell}}|z_{j_{1}},\ldots,z_{j_{m}};\theta),
    \end{split}
\end{equation*}
with $z_{i_{k}} = \xi^{-1}_{i_{k}}(x_{i_{k}}), k=1,\ldots,\ell$ and $z_{j_{k}} = \xi^{-1}_{j_{k}}(x_{j_{k}}), k=1,\ldots,m$.
Since $\mathcal{F}$ is stable by conditioning, then $F^{(d)}_{\mathbf{i}|\mathbf{j}}(\cdot;\theta)$ belongs to 
%$\xi \# \mathcal{F} \subset 
$\Xi \# \mathcal{F}$. The proof is similar for marginalization. 
\end{proof}
%\hfill $\square$

\vspace{1em}
This result can for instance be applied to the family of meta elliptical distributions introduced by Fang \cite{fang2002meta}, demonstrating that their conditional distributions also remain meta elliptical.
Note that we have only considered univariate transformations. However, for a fixed set of conditioning indices, it is worth mentioning that by following the same reasoning as in the proof, we can extend the property to blockwise bijective transformations. 
%This goes beyond the scope of the present paper.

\vspace{1em}
\begin{remark}\textbf{Stability by conditioning under linear forms.}
\label{rem:affine}
 In a similar manner, stability by conditioning under linear combinations of the components can also extend to mixtures and increasing transformations. In other words, if a family of distributions remains stable when conditioned through linear combinations of the components, the resulting families formed by mixing such distributions or transforming their marginal distributions will also exhibit stability. Notable examples of families possessing this property of stability include the multivariate Gaussian and multivariate Student's t-distributions.
\end{remark}
\vspace{1em}
\begin{remark}\textbf{Stability by conditioning of stochastic processes}

The distribution of a stochastic process is completely determined by its finite-dimensional distributions. This allows us to extend the aforementioned results for transdimensional families of multivariate distributions to stochastic processes.

Consider a stochastic process $\{X_{t}\}$ indexed by $t \in \mathcal{T}$ and valued in $\mathcal{X}$. Given its values $X_{t_{1}},\ldots,X_{t_{n}}$ at $n$ locations $t_{1},\ldots,t_{n} \in \mathcal{T}$, our goal is to determine its conditional distribution at $m$ new locations $t^{*}_{1}, \ldots,t^{*}_{m}$.

Assume the vector $\begin{pmatrix}
X_{\mathbf{t}} \\
X_{\mathbf{t}^{*}}
\end{pmatrix}$ , where $X_{\mathbf{t}} = (X_{t_1},\ldots,X_{t_{n}})^{t}$ and $X_{\mathbf{t^{*}}} = (X_{t^{*}_1},\ldots,X_{t^{*}_{m}})^{t}$, %= %\begin{pmatrix}
%X_{t_{1}} \\
%\vdots \\
%X_{t_{n}} \\
%X_{t^{*}_{1}} \\
%\vdots\\
%X_{t^{*}_{m}}
%\end{pmatrix}$
belongs to a transdimensional family of distributions stable by conditioning. Then by definition, the conditional distribution $X_{t^{*}_1},\ldots,X_{t^{*}_{m}}| X_{t_1}=x_1,\ldots,X_{t_{n}}=x_{n}$ remains within the same family of transdimensional distributions for every $x_{1},\ldots,x_{n} \in \mathcal{X}$. 

Utilizing the extension properties derived herein, any collection of functions $\{\xi_{t}\}, ~t \in \mathcal{T}, $ in $\mathcal{C}^{1}_\uparrow$, leads to a stochastic process $Z_{t} = \xi_{t}(X_{t})$ whose finite-dimensional distributions also belong to a transdimensional family stable under conditioning. This approach similarly applies to finite mixtures. An example of a mixture of transformed Gaussian Processes is provided example in Section \ref{sec:cond_process}, demonstrating how the conditional distribution can be computed analytically in such situations.

This allows to perform regression with stochastic processes beyond Gaussian processes. In practice, when selecting a family  of a stochastic process, it’s important to consider the data’s characteristics—such as tail properties, multi-modality, and marginal behaviors—to select an appropriate class of
distributions. With limited data, using a flexible model like mixtures can be challenging due
to the large number of parameters involved. If a mixture model appears beneficial,
one strategy could be to constrain the shapes of the covariance matrices, thereby reducing the number of parameters. Overall, we don't recommend using a broad model family to estimate numerous parameters when data is scarce.
    
\end{remark}
\vspace{1em}

\vspace{1em}

\section{Easy conditioning with fitted stable by conditioning distributions}
\label{sec::methods}

%As highlighted in the introduction, our focus is on estimating the conditional distributions of a (multivariate) response variable. This section delineates our methodology.

We now focus on the task of conditioning an absolutely continuous random vector $\mathbf{X}=(\mathbf{X}_{1}, \mathbf{X}_{2})$ taking values in $\prod_{k=1}^{d} \mathcal{X}_{k} = \mathbb{R}^{d}$ on the second group of variables $\mathbf{X}_{2}$. 
Our approach is to first fit a joint (easy to condition) distribution to the previously acquired sample $(\mathbf{x}_{1i},\mathbf{x}_{2i})_{i=1}^{n}$, and then, given any new observation $\mathbf{x}_{2}$ of $\mathbf{X}_{2}$, to leverage our results to estimate and/or sample from the conditional distribution of $\mathbf{X}_{1}$ given that $\mathbf{X}_{2}=\mathbf{x}_{2}$. Our generative procedure to estimate a conditional distribution thus consists in two main steps:

\vspace{0.5em}

\begin{itemize}
\item Fit an ``easy to condition'' joint distribution to a sample from $\mathbf{X}$

\vspace{0.5em}

\item perform conditioning of the estimated joint by leveraging analytical formulas.

\vspace{0.5em}

\end{itemize}

By using univariate transformations, we can model distributions with arbitrary (absolutely continuous) marginals. Specifically, if we consider the univariate transformations to be the marginal cumulative distribution functions, we obtain a space with uniform marginals and can subsequently apply copula theory, as detailed in the next section. In order to perform easy conditioning, we favor copula models corresponding to distributions stable by conditioning. 
Gaussian copulas, in particular, offer analytical formula for conditional distributions. However, they rely on latent multivariate Gaussian distributions, which are known to have notable limitations in terms of adequately model skewed, heavy-tailed, or multi-modal data. Alternative multivariate distributions, such as Student's t-distributions \cite{ding2016conditional} or skew-normal distributions \cite{azzalini2013skew}, offer more expressiveness and are stable by conditioning (see appendix for details). Resulting copula models can be used in our framework as they enable easy conditioning too. Yet, these distributions also fall short in scenarios requiring the modeling of multi-modal data and are confined to have marginal distributions of the same family along each axis. An important aspect of our the practical approach presented next is to allow capturing multi-modality by leveraging the extension of stability by conditioning to mixtures. The workhorse is to appeal to copulas of mixtures of multivariate distributions, where each individual component multivariate distribution possesses known conditional distributions. 

\subsection{Fitting joint distributions in terms of marginals and copulas}
\label{sec::copulas}
From a $d$-dimensional sample $\mathcal{D} =(x_{1i}, \ldots,x_{di})_{i=1}^{n} $ of independently drawn realizations, we want to estimate the distribution of the underlying random vector $\mathbf{X}$ taking values in $\mathbb{R}^{d}$. We use here a two-step procedure based on copulas. %, which are multivariate distribution functions with standard uniform margins.

Copulas provides a versatile framework for multivariate analysis. Following Sklar's theorem \cite{sklar1959fonctions}, for any distribution function $F$ with continuous margins $F_{j}, ~j=1,\ldots,d$, there exists a unique \textit{copula} $C$ such that

 \begin{equation*}
 \label{eq::sklar}
     F(x_{1},\ldots,x_{d}) = C(F_{1}(x_{1}),\ldots,F_{d}(x_{d})),~ \forall \textbf{x}=(x_{1},\ldots,x_{d}) \in \mathbb{R}^{d}.
 \end{equation*}
 
A multivariate distribution function can then be modeled separately by its marginals and its copula where the copula characterizes the dependence structure between the variables.

%Furthermore, if the marginals are continuous then $C$ is unique.

%A multivariate distribution function can be modeled separately by its marginals and its copula where the copula characterizes the dependence structure between components.

%Furthermore, we can obtain a similar result for the density of a multivariate distribution by introducing the notion of the density of a copula.

%\begin{prop}
%\label{prop:copula_density}
    %The density of a multivariate distribution $F$ is given by the following expression:
    %$$f(x_{1},\ldots,x_{d}) = c(F_{1}(x_{1}),\ldots,F_{d}(x_{d})) \prod_{i=1}^{d} f_{i}(x_{i}), ~\mathbf{x}=(x_{1},\ldots,x_{d}) \in \mathbb{R}^{d}$$ where $f_{i}$ is the density of the margin $F_{i}$ and $c$ is the density of the copula $C$ given by:
    %$$c(x_{1},\ldots,x_{d}) = \frac{ \partial C (u_{1},\ldots,u_{d})} {\partial u_{1} \ldots u_{d}}$$ where $u_{i} = F_{i}(x_{i}), ~i \in \{1,\ldots,d\}$.
%\end{prop}

\vspace{1em}

There exist several parametric families of copulas. In particular, for a given absolutely continuous multivariate distribution, we denote by “implicit copula” the underlying copula of the distribution. 

By “inverting” Sklar's theorem, if we assume that the marginal distributions  $F_{j}, ~j=1,\ldots,d$ are strictly increasing and continuous, we obtain that the copula can be written:
\begin{equation*}
    C(u_{1},\ldots,u_{d}) = F(F_{1}^{-1}(u_{1}), \ldots,F^{-1}_{d}(u_{d})),~ 0 \leq u_{1},\ldots,u_{d} \leq 1.
\end{equation*}
For example, the Gaussian copula, which is the copula implicit in a multivariate Gaussian distribution can be expressed:
$$C_{P}(\mathbf{u}) = \Phi_{P}(\Phi^{-1}(u_{1}),\ldots,\Phi^{-1}(u_{d})),$$
where $\Phi$ is the cdf of the univariate standard normal distribution and $\Phi_{P}$ is the cdf of the multivariate $\mathcal{N}_{d}(\mathbf{0},\mathbf{P})$ distribution with $\mathbf{P}$ a correlation matrix.

Differentiating the general formula with respect to $\mathbf{u}$, we get the density of the implicit copula by:

\begin{equation}
\label{eq::implicit_copula_density}
    c(u_{1},\ldots,u_{d}) = \frac{\partial^{d}}{\partial u_{1} \ldots \partial u_{d}} C(u_{1},\ldots,u_{d}) = \frac{f(z_{1},\ldots,z_{d})}{\prod_{i=1}^{d} f_{i}(z_{i})},
\end{equation}

%Assuming that the terms are non-zero, we can write:
%$$c(x_{1},\ldots,x_{d}) = \frac{f(x_{1},\ldots,x_{d})}{f_{1}(x_{1}) \times \ldots \times f_{p}(x_{p})} $$

where $z_{j} = F_{j}^{-1}(u_{j}), ~ j \in \{1,\ldots,d\}$, we say that $\mathbf{z}=(z_{1},\ldots,z_{d})$ is the quantile transform of $\mathbf{u}$, and $f$ and $f_i$ denote the densities associated with $F$ and $F_{i}$, respectively, for all $i \in {1, \ldots, d}$.

\vspace{1em}

%To estimate the joint distribution we use the decomposition of Sklar's theorem \ref{eq::sklar}. 
If the marginals are known, we can transform the observations $\mathbf{x}_{i}$ to pseudo-observations $\mathbf{u}_{i}$, with uniform marginals, by using the \textit{Probability Integral Transform (PIT)}, $u_{ji} = F_{j}(x_{ji})$,

$ ~\forall j \in \{1,\ldots,d\}$. And then fit a parametric copula $c(u_{1},\ldots,u_{d};\theta)$ to the pseudo observations by estimating the copula parameters $\theta$ with maximum-likelihood.

\vspace{1em}

The optimization problem is then:
\begin{equation*}
\label{eq::optimization}
    \hat{\theta} \in  \underset{\theta \in \Theta}{\mathrm{argmin}} \prod_{i=1}^{n} c(u_{1i},\ldots,u_{di};\theta).
\end{equation*}

However, in practice the marginal distributions are typically unknown, and one needs to estimate them from the data. Parametric or non-parametric methods can be used to estimate the marginal distributions. Parametric estimation of the marginals is used in the \textit{inference for margins (IFM)} method \cite{joe1996estimation}, while non-parametric estimation using the empirical distribution function is part of the \textit{pseudo likelihood estimator} used in \cite{genest1995semiparametric}, which has been proven to allow for asymptotical consistency in the estimation of dependence parameters.

%In this work, we will use a non-parametric method i.e kernel density estimation such that our algorithm doesn't need prior information about the marginals and is more generic. 

Once the marginals have been estimated, then we can plug-in the  “estimated pseudo
observations” , $\hat{u}_{ji} =\hat{F}_{j}(x_{ji})$ in the maximum-likelihood optimization problem.
%The procedure to estimate a joint multivariate distribution is summarized in Algorithm 1.

%\begin{algorithm}[H]
    %\label{alg::joint}
    %\caption{Pseudo Algorithm a learn the joint multivariate distribution}
    %\begin{algorithmic} 
    %\REQUIRE $\mathcal{D} = (x_{1i},\ldots,x_{di})_{i=1}^{n}$ a sample from $\mathbf{X} \in \mathbb{R}^{d}$.
    %\end{algorithmic}
    %\begin{itemize}
        %\item Estimate the marginal distributions $F_{j}, ~j \in \{1,\ldots,d\}$ by $\hat{F}_{j}$.
        
        %\item Transform each marginal to approximately uniform by PIT, $\hat{u}_{j} = \hat{F}_{j}(x_{j})$. 
        %\item Learn the copula function, which represents dependence structure by assuming a parametric form $c(\hat{u}_{1},\ldots,\hat{u}_{d},\theta)$.
        
    %\end{itemize}
%\end{algorithm}

\vspace{1em}

In our procedure, we assume that the joint distribution of the data is a meta distribution. A $d$-variate meta distribution is constructed by linking a $d$-variate implicit copula with $d$ arbitrary marginal cdfs. For example, a meta Gaussian distribution utilizes a Gaussian copula paired with arbitrary marginals, as a particular case of meta-elliptical distributions introduced by Fang \cite{fang2002meta}.

In this study, we work with parametric copulas implicit in a family of distributions stable by conditioning. Consequently, as derived from Theorem \ref{thm::transfo}, when considering strictly increasing marginal cdfs as a case of $\mathcal{C}_{\uparrow}^{1}(\mathbb{R},\mathbb{R)}$, the meta distributions considered also exhibit stability by conditioning.
For practical applications, our focus will be on copulas implicit in families of distributions for which analytical computation of conditional distributions is possible.%from a well known family. In particular, we choose families of distributions for which are stable by conditioning. For example, the conditional distributions of a meta Gaussian distributions are also meta Gaussian.

%We should mention that the stability of meta distributions corresponds to the stability of copulas. Our research focuses on parametric families of copulas for which the conditional copulas belongs to the same parametric family.

\subsection{Conditioning via latent space}

Once we have estimated the joint distribution, we estimate the conditional distribution (get conditional samples) given a new observation of the covariates. Since we assume that the data follow a "meta stable" distribution in order to perform easy conditioning, we are able to find a latent space in which we have stability by conditioning. In practice, we will focus on distributions for which there is an analytical formula for the conditional distributions.

%Let introduce the notion of “implicit copula” which refers to the copula implicit in the multivariate distribution of a continuous random vector.
%By "inverting" Sklar's theorem, if we assume that the marginal distributions  $F_{j}, ~j=1,\ldots,d$ are strictly increasing, we obtain that the copula can be written:
%\begin{equation}
    %C(u_{1},\ldots,u_{d}) = F(F_{1}^{-1}(u_{1}), \ldots,F^{-1}_{d}(u_{d})),~ 0 \leq u_{1},\ldots,u_{d} \leq 1
%\end{equation}
%Differentiating with respect to $\mathbf{u}$, we get the density of the implicit copula by:

%\begin{equation}
%\label{eq::implicit_copula_density}
    %c(u_{1},\ldots,u_{d}) = \frac{\partial^{d}}{\partial u_{1} \ldots \partial u_{d}} C(u_{1},\ldots,u_{d}) = \frac{f(z_{1},\ldots,z_{d})}{\prod_{i=1}^{d} f_{i}(z_{i})}
%\end{equation}

%Assuming that the terms are non-zero, we can write:
%$$c(x_{1},\ldots,x_{d}) = \frac{f(x_{1},\ldots,x_{d})}{f_{1}(x_{1}) \times \ldots \times f_{p}(x_{p})} $$

%where $z_{j} = F_{j}^{-1}(u_{j}), ~ j \in \{1,\ldots,d\}$, we say that $\mathbf{z}=(z_{1},\ldots,z_{d})$ is the quantile transform of $\mathbf{u}$.

\vspace{1em}

Copulas transport the problem from the original space $\mathcal{X}$ to uniform space $\mathcal{U}=[0,1]^{d}$. 
%because it is usually easier to capture multivariate dependence using $C$ on $[0,1]^{d}$. 
We go one step further by transforming the marginals to a latent space $\mathcal{Z}$ \cite{smith2023implicit}, in which the marginals are those of an implicit multivariate distribution driving the copula. The two consecutive transformations are illustrated in the following scheme:

\begin{figure}[H]
    \centering
    \includegraphics[scale=1]{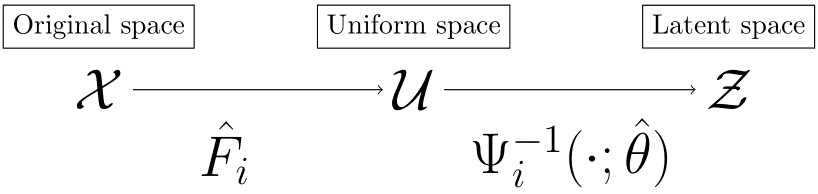}
    \label{fig:latent_space}
\end{figure}

\noindent
where $\Psi_{i}(\cdot;\hat{\theta})$ represents the cumulative distribution function of the $i$-th marginal of an implicit multivariate distribution linked to the estimated copula parameters $\hat{\theta}$. In the latent space $\mathcal{Z}$, the samples follow this implicit multivariate distribution.
%\subsection{Conditioning in the latent space}
%The goal is to have as much flexibility as possible in the joint distribution while enabling efficient and exact conditioning. In order to be able to perform exact conditioning in the latent space, we will use implicit copulas of multivariate distributions enjoying the EC property. 

%If we assume the copula to be the implicit copula of a multivariate distribution, then in the latent space, the samples will be distributed following this multivariate distribution.
By our assumption of “meta stable” distribution, we can use analytical formulas for conditioning in the latent space. Finally we transform back to the original space to get an estimate of the conditional distribution with the right marginals.

The full pseudo-algorithm to estimate a conditional distribution function is presented next:

%\vspace{1em}

\begin{algorithm}[H]
    \label{alg::joint}
    \caption{Estimation of conditional CDF via latent space}
    \begin{algorithmic} 
    \REQUIRE $\mathcal{D} = (x_{1i},\ldots,x_{di})_{i=1}^{n}$ a sample from $\mathbf{X} \in \mathbb{R}^{d}$, $\ell$ the dimension of the subvector of $\mathbf{X}$ which conditional distribution is sought, and $\mathbf{x}^{(new)}_{\ell + 1:d} = (x^{(new)}_{\ell+1},\ldots,x^{(new)}_{d})$ a realization of the conditioning variables.
    \end{algorithmic}
    \begin{itemize}
       \item Estimate the marginal distributions $F_{j}, ~j \in \{1,\ldots,d\}$ by $\hat{F}_{j}$.
        
        \item Transform each marginal to approximately uniform by PIT, $\hat{u}_{j} = \hat{F}_{j}(x_{j})$. 
        
        \item Learn the copula function, which represents dependence structure by assuming a parametric form $c(\hat{u}_{1},\ldots,\hat{u}_{d},\theta)$. This is the implicit copula of a multivariate distribution with marginals $\Psi_{1},\ldots,\Psi_{d}$.
        
        \item Transform the new observed point to the latent space $\mathcal{Z}$ by applying the transform  $$\mathbf{z}^{(new)}_{\ell+1 : d}=(z^{(new)}_{\ell+1},\ldots,z^{(new)}_{d}) = (\Psi^{-1}_{\ell+1}\circ \hat{F}_{\ell+1}(x_{\ell+1}^{(new)}),\ldots, \Psi^{-1}_{d} \circ \hat{F}_{d}(x_{d}^{(new)}))$$
        \item Perform conditioning in the latent space by using the analytical formulas to get an estimate of the cdf: $$\hat{F}_{\mathbf{Z}_{1}|\mathbf{Z}_{2}= z^{(new)}_{\ell +1 :d}}(z
        _{1}^{(new)},\ldots,z_{\ell}^{(new)}|z^{(new)}_{\ell+1:d})$$

        \item Transform back to get an estimate of the cdf in the original space by:
        $$\hat{F}_{\mathbf{X}_{1}|\mathbf{X}_{2}
        %=\mathbf{x}^{(new)}_{\ell + 1:d} 
        }\left(x_{1},\ldots,x_{\ell}\big|\mathbf{x}_{\ell+1:d}^{(new)}\right) = \hat{F}_{\mathbf{Z}_{1}|\mathbf{Z}_{2}
        %= \mathbf{z}^{(new)}_{\ell+1:d}
        }
        \left(\hat{F}_{1}^{-1} \circ \Psi_{1} (z_{1}^{(new)}),\ldots,\hat{F}_{\ell}^{-1} \circ\Psi_{\ell}(z_{\ell}^{(new)}) \big| \mathbf{z}_{\ell+1:d}^{(new)} \right)$$
        
    \end{itemize}
\end{algorithm}

\begin{figure}[H]
    \centering
    \includegraphics[scale=0.65]{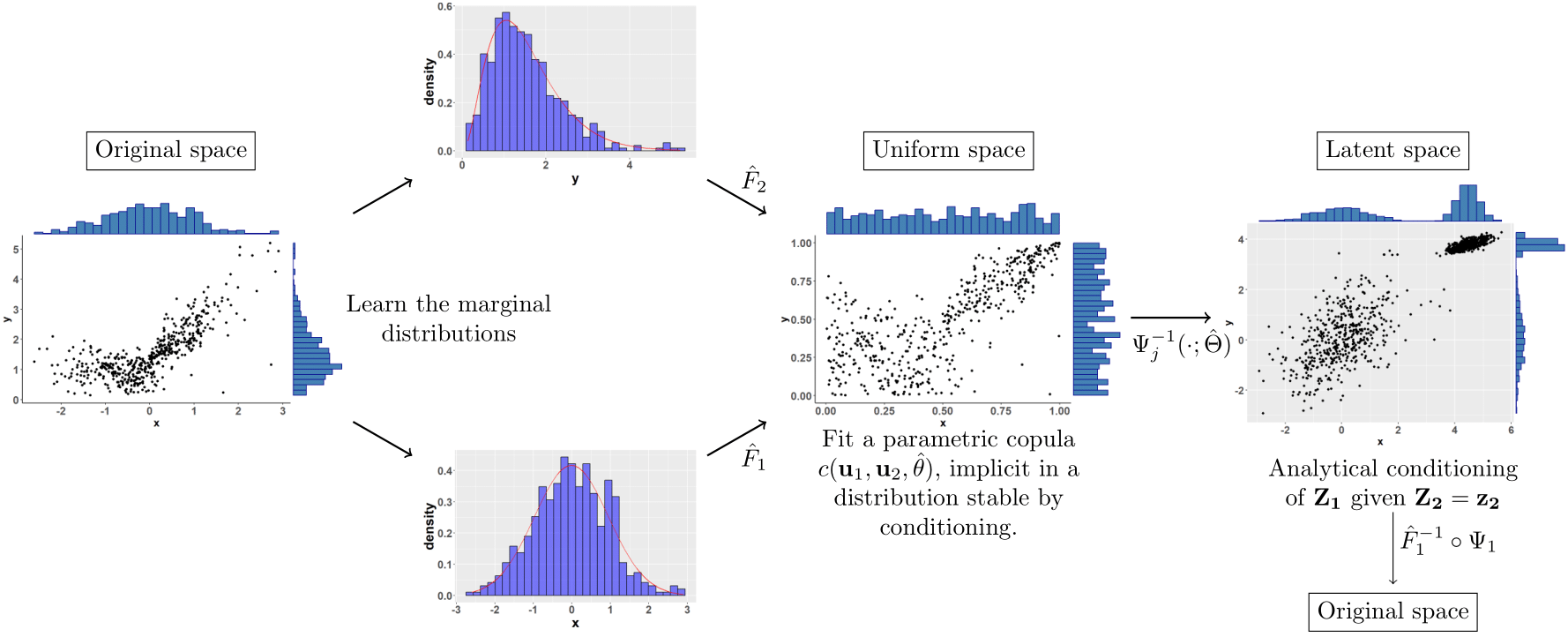}
    \caption{Workflow of the conditioning algorithm on a $2$-dimensional example.}
    \label{fig:workflow}
\end{figure}

The algorithm above can be easily adapted to sample from the conditional distribution. 
The full workflow for estimating conditional distributions is illustrated on a $2$-dimensional example in figure \ref{fig:workflow}.

\subsection{Implementation of the method: the example of the Gaussian Mixture Copula Model}
\label{sec:gmcm}

Our method's versatility stems from its ability to choose any dependence structure among families of copulas  stable by conditioning, which includes mixtures, and to model the joint distribution with arbitrary marginals. The primary challenge in our procedure lies in fitting a copula in the uniform space. This involves balancing the complexity of the dependence structure against the precision with which we can estimate the copula parameters. For this paper, we have opted to use a Gaussian mixture copula with a limited number of components. Using further dependence structures such as Student's $t$-mixture copulas or Skew-Normal mixture copulas represent exciting avenues for future research.
%Our method is very versatile since it allows to choose any dependence structures among the multivariate distributions with EC property (which includes mixtures), and arbitrary marginals, to model the joint distribution. In our procedure, the most challenging step is to fit a copula in the uniform space. There is a trade-off between allowing for a complex dependence structure and being able to fit the copula parameters precisely. In this paper, we have decided to use the Gaussian mixture copula with a limited number of components, but other dependence structures such as Student $t$-mixture copulas or Skew-Normal mixture copulas can be explored in future works.

%\subsection{\textcolor{red}{Definition}}

The Gaussian mixture copula \cite{tewari2011parametric} is the copula implicit in a Gaussian mixture model. It allows to capture multi-modality in the dependence structure. The density of a Gaussian mixture model (GMM) with $K$ components is given by:
\begin{equation*}
    \psi(x_{1},\ldots,x_{d}, \theta) = \sum_{k=1}^{K} \alpha_{k} \phi(x_{1},\ldots,x_{d};\mu_{k},\Sigma_{k}),
\end{equation*}
with $\alpha_{k} \geq 0,~\forall k \in \{1,\ldots,K\}$ such that $\sum_{k=1}^{K} \alpha_{k}=1$, the mixing weights and $\mu_{k},~\Sigma_{k}$ the mean and covariance matrix of the $k$-th component. $\theta = \{ \alpha_{i},\mu_{i},\Sigma_{i}\}_{i=1}^{K}$ is the set of all parameters and $\phi$ the pdf of a Gaussian distribution.

Then from the equation \ref{eq::implicit_copula_density}, we have that the density of a Gaussian mixture copula with parameters $\theta$ is given by:
    \begin{equation*}
    \label{eq::density_gmcm}
        c_{gmc}(u_{1},\ldots,u_{d};\theta)=\frac{\psi(\Psi^{-1}_{1}(u_{1}),\ldots,\Psi^{-1}_{d}(u_{d}))}{\prod_{i=1}^{d} \psi_{i}(\Psi_{i}^{-1}(u_{i}))}, ~ 0 \leq u_{1},\ldots,u_{d} \leq 1,
    \end{equation*}
    where $\psi_{j}$ and $\Psi_{j}^{-1}$ denote the marginal density and inverse distribution function of the GMM along the $j$-th dimension .

Similarly, we can define the Student $t$-mixture copula \cite{massing2021clustering} or the unified Skew-Normal mixture copula as other implicit copulas.

Estimating parameters of a GMCM is a challenging task for several reasons as it has been discussed in \cite{tewari2023estimation}, we provide details about fitting of a GMCM and compare several methods in the appendix \ref{sec:gmcm}.

\subsection{Computational cost vs methodological benefits}
\label{sec:compute_time}

In this section, we would like to emphasize that while our method incurs higher computational costs compared to some other conditional distribution estimation techniques, it also offers significant advantages. 

The primary computational costs stem from estimating parameters of a Gaussian Mixture Copula Model (GMCM) to fit the joint distribution and from evaluating the quantile function of Gaussian mixture models during the conditioning step. Estimating the parameters of a GMCM is a challenging task, and further details on the fitting procedure are provided in Appendix~\ref{sec:details_gmcm}. Although parameter estimation for GMCMs remains an active area of research, we employ the most efficient methods currently available to the best of our knowledge. Nevertheless, further methodological developments may improve the computational efficiency of the fitting procedure. However, this is beyond the scope of the present work and is therefore left for future research.

To evaluate the quantile function, essential for converting back samples to the original space post-conditioning, we use the Chandrupatla root-finding algorithm \cite{chandrupatla1997new}. Our algorithm needs to evaluate quantiles for each generated sample, which quickly becomes computationally demanding.

%\textcolor{red}{While there are potential optimizations for our method to reduce computational time, these are beyond the scope of this paper and are left for future exploration.}

In contrast, methods like distributional random forests \cite{cevid2022distributional}, used in our experiments (Section \ref{sec:experiments}), generate conditional samples significantly faster and demonstrate slightly better performance compared to our copula-based approaches.

Despite this, our method offers several advantages compared to distributional random forests. Notably, it provides analytical formulas for conditional distributions, enabling direct computation of quantities of interest such as conditional moments, probabilities, or other functionals, and application to the conditioning of stochastic processes, as shown in Section \ref{sec:cond_process}. 

Our approach also ensures
coherence across conditioning queries, which is typically not guaranteed when relying on separately fitted machine-learning models for different conditioning queries. Moreover, explicit probabilistic models make the necessary assumptions about tail behaviour visible and propagate them coherently under conditioning, including for conditioning events or scenarios with little or no direct empirical support.

Additionally, the generative nature of our method—fitting the joint distribution and then conditioning—differs from discriminative methods like random forests. This allows to apply our method to a wider range of tasks such as synthetic data generation. Another advantage is the amortization, meaning that fitting a single joint distribution is sufficient to generate samples from any conditional distributions. This is particularly useful for imputation tasks (see Section \ref{sec:imputation}) where different variables may be missing across observations.

Finally, our approach allows for conditioning with respect to linear combinations of the components, as detailed in Remark \ref{rem:affine}, offering valuable flexibility in numerous applications such as linear inverse problems \cite{stuart2010inverse, travelletti2023uncertainty}. To our knowledge, this capability is not directly achievable with random forest-based methods.

\section{Numerical experiments}
\label{sec:experiments}
\subsection{Conditioning multivariate distributions}

In this section, we address the challenge of conditioning multivariate distributions. Specifically, for a given dataset, our goal is to sample from the conditional distribution of the missing variables based on the observed variables. We achieve this using the following approach.

We split the considered data sets randomly into a training set containing $ 80 \%$ of the observations and a test set containing the remaining $20 \% $ of the observations. We use the training set to learn the joint multivariate distribution and we sample $1000$ points from the estimated conditional distribution at every test points.

We then evaluate the performance of our approach comparing the conditional samples to the observed values using scoring rules. In the univariate case, we use the Continuous Ranked Probability Score (CRPS) and the logarithmic score (LogS). For multivariate outcomes, we employ the Energy score (ES) and the Variogram score (VS) \cite{jordan2017evaluating}, with order $r=0.5$. In particular, we will examine the mean value of the various scores across all test points, averaged over multiple train/test splits.

We denote the method presented in Section \ref{sec:gmcm} as GMCM. The marginal distributions are learned parametrically by assuming a Gaussian mixture model, with the number of components chosen automatically between 1 and 10 based on the Akaike Information Criterion (AIC). However, the number of components in the Gaussian mixture copula must be manually specified. We ideally recommend to use the understanding of the data to determine the number of mixture components. For instance, if the dataset
represents a population with two distinct groups, starting with two components could be sensible. When such knowledge is unavailable, a clustering
method might help determine the number of components. 
To compute the quantile function of the marginal distributions needed in Algorithm \ref{alg::joint}, we employ the numerical Chandrupatla root-finding algorithm \cite{chandrupatla1997new}.

We compare our approach to a method using a Gaussian copula, referred to as GC. This method operates similarly to ours in learning the marginal distributions but assumes a Gaussian copula instead of a Gaussian mixture copula. The conditioning is performed in a latent Gaussian space.

Given the complexity of fitting a GMCM in the uniform space, an alternative approach for conditional sampling involves fitting a GMM in a transformed space using the EM algorithm, which boasts theoretical convergence guarantees. Specifically, we can
transform each marginal distribution to an approximate standard normal distribution. Following this, we can perform conditioning
within this space and subsequently transform the samples back to the original space. We
will refer to this method as Transformed Gaussian Mixture Model (TGMM).

%Finally, we benchmark our approach against a Conditional Kernel Density Estimation (CKDE) method, where the joint density is estimated using KDE, and samples are drawn from the conditional density by filtering based on localization conditions, see details in the appendix.
Finally, we benchmark our approach against a distributional random forest (DRF) method developped by Cevid and al. \cite{cevid2022distributional} for distributional regression tasks.

Note that, given $d$ the dimension of the joint distribution, the number of parameters for the GC method is: $\frac{d \times d-1}{2} +d$. If we denote $k$ the number of components of the GMCM and TGMM methods, then the number of parameters is multiplied by $k$ for these two methods.
\subsubsection{Synthetic examples}

%\vspace{1em}

 %\textbf{Simulation scenarios:}
 
 %\vspace{1em}
 
We consider two simulation scenarios  focused on sampling from a univariate conditional distribution:

\vspace{0.5em}
\begin{itemize}
    \item \textbf{Gaussian mixture model}:

    \vspace{0.5em}
    
The data points are generated from a Gaussian mixture model in dimension two, with known parameters. 

\noindent
Here, we take $(\mathbf{X}_{1},\mathbf{X}_{2}) \sim \text{GMM}(\{\alpha_{i},\mu_{i},\sigma_{i}\}_{i=1}^{2})$ with $\alpha_{1} = 0.3, ~\alpha_{2} = 0.7$ and 

\vspace{1em}
$\mu_{1} = \begin{pmatrix}
    4 \\
    2
\end{pmatrix}, ~\mu_{2} =  \begin{pmatrix}
    -2 \\
    1
\end{pmatrix}$ and $\Sigma_{1} = \begin{pmatrix}
    2 & 1 \\
    1 & 1
\end{pmatrix}, ~ \Sigma_{2} = \begin{pmatrix}
    1 & 0.5 \\
    0.5 & 1
\end{pmatrix}$.

\vspace{1em}

\noindent
%Samples from this multivariate distribution are represented %in the supplementary material
%on figure \ref{fig::descriptive_gmm}.

In this case, we know analytically the form of the conditional distribution:

\begin{equation*}
    X_{1} \mid X_{2} = x_{2} \sim \Tilde{\alpha}_{1}(x_{2}) \mathcal{N}(4 + (x_{2}-2), 1) + \Tilde{\alpha}_{2}(x_{2}) \mathcal{N}\left(\frac{x_{2} - 1}{2} - 2, 0.75\right)
\end{equation*}
with $\Tilde{\alpha}_{1}(x_{2}) = \frac{0.3 \mathcal{N}(x_{2};2,1)}{0.3 \mathcal{N}(x_{2};2,1) + 0.7 \mathcal{N}(x_{2};1,1)} $ and $\Tilde{\alpha}_{2}(x_{2}) = \frac{0.7 \mathcal{N}(x_{2};1,1)}{0.3 \mathcal{N}(x_{2};2,1) + 0.7 \mathcal{N}(x_{2};1,1)}$
.

\vspace{0.5em}

In this test case, the data follows a Gaussian mixture dependence structure, and we know the specific number of components. 
%As a result, we anticipate that our method GMCM, will perform very well.

\vspace{0.5em}

\item \textbf{Meta Gaussian mixture model:}

\vspace{0.5em}

We now consider another example with a univariate response, where the data has been simulated from a $2$-dimensional meta-Gaussian mixture model. 
%i,e, with a copula corresponding to that of a Gaussian mixture model and arbitrarily chosen 
Here the marginal distributions are chosen standard normal, and the dependency structure is driven by a Gaussian mixture copula model with the parameters  $\alpha_{1}= \alpha_{2}=0.5$ and 

\vspace{1em}
$\mu_{1} = \begin{pmatrix}
    2 \\
    2
\end{pmatrix}, ~\mu_{2} =  \begin{pmatrix}
    -2 \\
    -2
\end{pmatrix}$ and $\Sigma_{1} = \begin{pmatrix}
    1 & 0.5 \\
    0.5 & 1
\end{pmatrix}, ~ \Sigma_{2} = \begin{pmatrix}
    1 & 0.5 \\
    0.5 & 1
\end{pmatrix}$.

\vspace{1em}

%The marginals are chosen to be standard normal distributed. 
%An illustration of this multivariate distribution can be found in %the supplementary material. 
%figure \ref{fig:copulogram}.

\vspace{0.5em}

\end{itemize}

\vspace{1em}
\textbf{Results}: 

\vspace{1em}

Since we have considered synthetic examples where the response is univariate and the cumulative distribution function (cdf) of the conditional distribution has an analytical form, we can visually compare the cdf of the true conditional distribution with the cdf of the estimated conditional distribution produced by our method, at various new locations. Additionally, we compare the various methods using univariate scoring rules.

\begin{figure}
    \centering
    \includegraphics[scale=0.7]{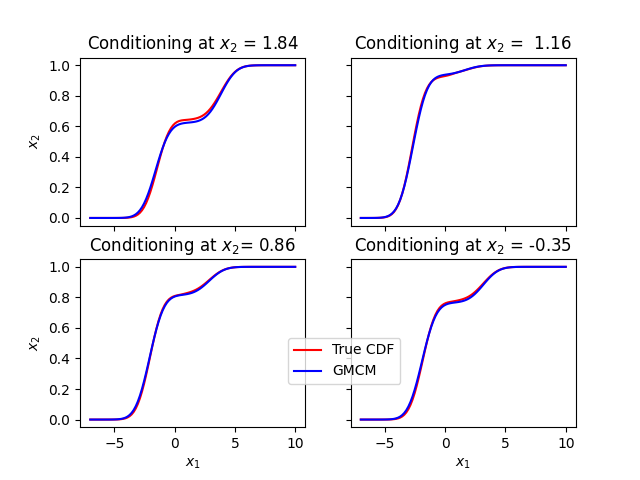}
    \caption{Comparison of true conditional cdf and estimated conditional cdf at various locations. The data follows a Gaussian mixture model distribution.}
    \label{fig:analytical_example}
\end{figure}

\begin{figure}
    \centering
    \includegraphics[scale=0.7]{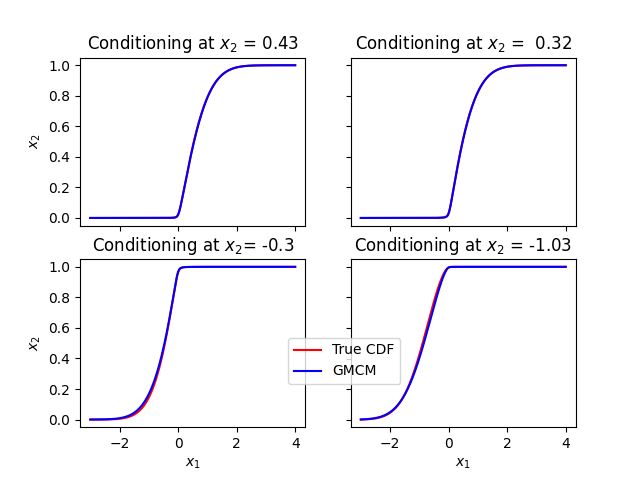}
    \caption{Comparison of true conditional cdf and estimated conditional cdf at various locations. The data follows a meta-Gaussian mixture model distribution.}
    \label{fig:meta_cdfs}
\end{figure}

In figure \ref{fig:analytical_example} and \ref{fig:meta_cdfs}, we represent the true cdf of the conditional distribution and the estimated cdf by our method using GMCM, at various conditioning locations. The true cdf is depicted with a blue line and the cdf estimated by our method by a red line.

%\subsubsection{Gaussian mixture model}
%Let us consider an example in $2$-dimensions where data points arises from a Gaussian mixture model with known parameters. 

Based on these figures, it appears that our algorithm accurately recovers the conditional distributions when the joint distribution is generated from a Gaussian mixture model or from a meta Gaussian mixture model, when the number of components is known.

Additionally, we compare the various methods using univariate scoring rules.

  \begin{center}
\begin{table}[H]%
\caption{Comparison of various methods based on average scores over $10$ train/test splits for the $2$ simulation scenarios.}
\vspace{1em}
\label{table}
  \label{table:cut-off_0.01} 
\begin{tabular*}{400pt}{@{\extracolsep\fill}lcccccc@{\extracolsep\fill}}
 & \multicolumn{2}{c}{\textbf{GMM}} & \multicolumn{2}{c}{\textbf{Meta GMM}} \\  
 & \textbf{CRPS} & \textbf{LogS} & \textbf{CRPS} & \textbf{LogS} \\  
 \toprule 
 \textbf{GC}   & 1.249 $\pm$ 0.067  & 1.894 $\pm$ 0.051  & 0.317 $\pm$ 0.016 &  0.891 $\pm$ 0.052 \\ 
\textbf{GMCM} &  1.248 $\pm$ 0.066 & 1.867 $\pm$ 0.040  & 0.288 $\pm$ 0.017 & 0.680 $\pm$ 0.058 \\ 
%\textbf{GMCM Cond Transform} & 1.738   &  & 0.580 &  \\
\textbf{TGMM}& 1.251 $\pm$  0.074 & 1.890 $\pm $ 0.047   & 0.299 $\pm$ 0.018 & 0.776 $\pm $ 0.058\\
\textbf{DRF} & 1.273 $\pm $ 0.076  & 1.915 $\pm$ 0.072 &  0.293 $\pm$ 0.019 &  0.776 $\pm$  0.110\\
\bottomrule
\end{tabular*}
\label{table_meta_gmm}

\end{table}
\end{center}

The results indicate that our method using GMCM seems to outperforms the other competitors for conditional sampling on simple examples. 
% Notice that a convenient feature of the considered approach is that it provides estimates of the cdf $\hat{F}_{\mathbf{X_{1}}|\mathbf{X_{2}} = \mathbf{x_{2}}}(\mathbf{x_{1}})$ which are smooth both in $\mathbf{x_{2}}$ and in $\mathbf{x_{1}}$. 
%However, we need to fix the number of mixtures components beforehand, which would require some prior knowledge about the data. One potential solution is to select a large number of mixture components to enhance the modeling capacity of our model, thereby capturing more complex dependence structures. Nonetheless, this approach introduces greater challenges in parameter estimation.
%Consequently, the advantage gained from a more comprehensive modeling of the joint distribution is offset by the reduced accuracy in parameter estimation.
In the next section, we turn to real datasets with multivariate outputs. 

\subsubsection{Real world examples}
% To evaluate the quality of the conditional samples produced by our algorithm, we will employ multivariate scoring rules. To compare the different methods, we will examine the mean scores over the test set, averaged across multiple train/test splits. As competitors, we will again consider a Gaussian copula model (GC) and the TGMM method.
\vspace{1em}

\textbf{Wine dataset:}

\vspace{1em}

This dataset already presented in the introduction contains the amount of $13$ constituents found in each of the three types of wines and consists of $178$ observations. We set the number of components in our GMCM to 3. For our analysis, we work with a $5$-dimensional problem where we try to predict the $3$ first variables given the $2$ remaining ones, based on training data consisting of joint observations of the $5$ considered variables. 
%A graphical representation of this dataset can be found %in the supplementary material 
%in figure \ref{fig:wine_descriptive}.

%We compare our method with the two competitors described above.
In Table \ref{table_wine}, we present the mean and standard deviation results of multivariate scoring rules across 10 train/test splits. The results indicate that GMCM performs slightly better than GC and TGMM in terms of the Energy Score (ES), a scoring rule extending CRPS to multivariate settings. However, it does not outperform them in the variogram score (VS). Notably, the DRF benchmark appears to outperform all other methods in this example. Given that the performance differences among the models are within one standard deviation, we cannot assert a significant difference in their performance.% This could be attributed to the modeling of marginal distributions, for which we assume a Gaussian mixture form in our methodology, and seems not to be adequate in this example.
  \begin{center}
\begin{table}[H]%
\caption{Comparison of various methods based on average scores over the test set for the Wine dataset.}
\vspace{1em}
\label{table}
  \label{table:cut-off_0.01} 
\begin{tabular*}{400pt}{@{\extracolsep\fill}lccc@{\extracolsep\fill}}
 & \textbf{ES} & \textbf{VS} %& %\textcolor{red}{Time (s)} 
 \\ 
 \toprule
\textbf{GC} & 0.789 $\pm$ 0.044 & 0.262 $\pm$ 0.034 %& \textcolor{red}{138.3}
\\ 
\textbf{GMCM} & 0.787 $\pm$ 0.063 & 0.268 $\pm$ 0.115  %& \textcolor{red}{200.7}
\\ 
%\textbf{GMCM cond transport} & 1.2531 & \\
\textbf{TGMM} & 0.851 $\pm$ 0.183 & 0.305 $\pm$ 0.041 %& \textcolor{red}{142.4}
\\
\textbf{DRF} & 0.774 $\pm$ 0.046 & 0.262 $\pm$ 0.037 %& \textcolor{red}{6.9}
\\
\bottomrule
\end{tabular*}
\label{table_wine}
\vspace{1em}

\end{table}
\end{center}

\textbf{Breast Cancer data Wisconsin:}

\vspace{1em}

\label{sec::wisconsin}

The Breast Cancer Wisconsin dataset sourced from the UCI Machine Learning Repository, utilizes digitized images from fine needle aspirates (FNA) of breast masses. It comprises ten features derived from \(569\) patients. For each feature, the dataset provides the mean value, extreme value (average of the three largest values), and standard error, resulting in \(30\) continuous variables. The dataset includes two types of diagnoses: benign (\(352\) patients) and malignant (\(212\) patients), suggesting the presence of two distinct clusters. For this experiment, we focus on a subset of four variables: perimeter standard error (PSE), extreme smoothness (ES), extreme concavity (EC), and extreme concave points (ECP).

We aim to use probabilistic methods to predict the values of EC and ECP based on observations of PSE and ES.

%A graphical representation of the data-set can be found %in the supplementary material.
%in figure \ref{fig:cancer_descriptive}. 

Given that the dataset appears to contain two clusters corresponding to malignant and benign diagnoses, we will set the number of components in our GMCM to 2. %We sample $n=1000$ points for the conditional distribution at every test points. 
The results in Table \ref{table_breast} indicate that the GMCM method performs effectively in this scenario and appears to outperform the other methods. However, similar to the previous experiment, the performance differences among the models are minimal, making it challenging to demonstrate a significant difference in their effectiveness.

  \begin{center}
\begin{table}[H]%
\caption{{Comparison of various methods based on mean $\pm$ std of scoring rules over $10$ train/test splits for the Wisconsin Breast Cancer dataset}}
\vspace{1em}
\label{table}
  \label{table:cut-off_0.01} 
\begin{tabular*}{400pt}{@{\extracolsep\fill}lccc@{\extracolsep\fill}}
 & \textbf{ES} & \textbf{VS} \\ 
 \toprule
\textbf{GC} & 0.744 $\pm$ 0.089 & 0.329 $\pm$ 0.080\\ 
\textbf{GMCM} &  0.717 $\pm$ 0.093 & 0.306 $\pm$ 0.080  \\
%\textbf{GMCM Cond Transport} & \textbf{2.275} & \\
\textbf{TGMM} & 0.722 $\pm$ 0.101 &  0.316
$\pm$ 0.092
\\ \textbf{DRF} & 0.734 $\pm$ 0.096 &  0.316 $\pm$  0.078\\
\bottomrule
\end{tabular*}
\label{table_breast}
%\vspace{1em}
\end{table}
\end{center}

\vspace{1em}

\textbf{Multiscale imaging of the human hippocampus data:}

\vspace{1em}

In a study examining the three-dimensional (3D) cytoarchitecture of the human hippocampus in both neuropathologically healthy individuals and those with Alzheimer's disease (AD), Eckermann et al. \cite{eckermann2021three} utilized phase-contrast X-ray computed tomography to analyze postmortem human tissue punch biopsies. Tissue samples from 20 individuals were scanned and evaluated using an automated measurement and analysis approach.

Following pre-processing and segmentation, each individual was represented by a point cloud in a five-dimensional space of structural parameters, with each point corresponding to a cell. For each cell, five variables were measured: median electron density, heterogeneity, volume, sphericity, and the number of neighbors within a radius of $13.5 \mu m$. The dataset is publicly available on Zenodo \cite{eckermann2022data}. In the example below, we randomly select $5$ patients and $1,000$ cells, aiming to predict heterogeneity, sphericity, and the number of neighbors within the specified radius based on electron density and volume. We report the mean and standard deviation of our probabilistic predictions over $5$ train/test splits.%In the example below, we examine a single patient with $1,000$ cells, aiming to predict heterogeneity, sphericity, and the number of neighbors within the specified radius based on electron density and volume. We report the mean and standard deviation of our probabilistic predictions over $5$ train/test splits.
%\begin{center}
%\begin{table}[H]%
%\caption{Comparison of various methods based on mean $\pm$ std of scoring rules over $5$ train/test splits.}
%\vspace{1em}
%\label{table}
  %\label{table:cut-off_0.01} 
%\begin{tabular*}{400pt}{@{\extracolsep\fill}lccc@{\extracolsep\fill}}
 %& \textbf{ES} & \textbf{VS} \\ 
 %\toprule
%\textbf{GC} & 56.49 $\pm$ 1.27 & 19.71 $\pm$ 1.37 \\ 
%\textbf{GMCM} & 51.46 $\pm$ 1.04 & 17.27 $\pm$ 1.57\\ 
%\textbf{GMCM cond transport} & 1.2531 & \\
%\textbf{TGMM} & 50.93 $\pm$ 0.87 & 17.02 $\pm$ 1.47\\
%\textbf{DRF} & 58.88 $\pm$  3.39 &  22.31 $\pm$ 2.54\\
%\bottomrule
%\end{tabular*}
%\label{table_eckermann}
%\vspace{1em}

%\end{table}
%\end{center}

\begin{center}
\begin{table}[H]%
\caption{Comparison of various methods based on mean $\pm$ std of scoring rules over $5$ train/test splits for $5$ randomly selected patients.}
\vspace{1em}
\label{table}
  \label{table:cut-off_0.01} 
\begin{tabular*}{400pt}{@{\extracolsep\fill}lccc@{\extracolsep\fill}}
 & \textbf{ES} & \textbf{VS} \\ 
 \toprule
\textbf{GC} & 83.85 $\pm$ 7.78 & 34.86 $\pm$ 7.36 \\ 
\textbf{GMCM} & 73.02 $\pm$ 3.87 & 27.51 $\pm$ 2.24\\ 
%\textbf{GMCM cond transport} & 1.2531 & \\
\textbf{TGMM} & 76.65 $\pm$ 7.37 & 31.31 $\pm$ 7.43\\
\textbf{DRF} & 71.52 $\pm$  5.52 &  28.30 $\pm$ 4.25\\
\bottomrule
\end{tabular*}
\label{table_eckermann}
\vspace{1em}

\end{table}
\end{center}
The results in Table \ref{table_eckermann} show that GMCM and DRF outperform GC and TGMM for this task. Additionally, the variability in the results for GC and TGMM is more pronounced than for GMCM.
%The results in Table \ref{table_eckermann} indicate that methods using mixtures of distributions, specifically GMCM and TGMM, outperform DRF and GC for this task. This suggests an underlying clustering structure in the dataset.

\begin{remark}
    In these examples, our GMCM approach requires significantly more computational time than DRF to achieve similar performances. For conditional sampling tasks, we recommend using DRF due to its flexibility and speed. However, as discussed in Section \ref{sec:compute_time} and demonstrated below, our method offers distinct advantages for a wide range of other applications.
\end{remark}
\subsection{Easy conditioning of stochastic processes}
\label{sec:cond_process}

In this section, we provide an example of a stochastic process constructed through mixtures and transformations of a Gaussian process, which is easy to condition. We demonstrate how to draw sample from its conditional distribution at a new location.

Consider two real-valued independent Gaussian processes, $(Z^{(1)}_{t})_{t \in \mathcal{T}}$ and $(Z^{(2)}_{t})_{t \in \mathcal{T}}$ defined on a set $\mathcal{T}$. %with mean zero and RBF kernels with different bandwidth. 
Define the transformed processes as $W^{(1)}_{t} = g_{t}(Z^{(1)}_{t}), ~t \in \mathcal{T}$ and $W^{(2)}_{t} = g_{t}(Z^{(2)}_{t}), ~t \in \mathcal{T}$, where $g_{t} : x \rightarrow \exp{(tx)}$ is a strictly increasing function. Then, consider the mixture process with $p \in (0,1)$: \[
W_{t} = 
\begin{cases} 
W^{(1)}_{t} & \text{with probability } p \\ 
W^{(2)}_{t} & \text{with probability } 1-p 
\end{cases}
\]

Having observed the stochastic process $W_{t}$ at $n$ locations $t_{1},\ldots,t_{n} \neq 0$, we would like to compute the conditional distribution of this stochastic process at $m$ new locations $t^{*}_{1},\ldots,t^{*}_{m}  \neq 0$ i.e the conditional distribution of $W_{t^{*}_{1}},\ldots,W_{t^{*}_{m}}|W_{t_{1}},\ldots,W_{t_{n}}$.
By properties of Gaussian processes, we have that for $i \in \{1,2\}$:
$$Z^{(i)}_{t^{*}_{1}},\ldots,Z^{(i)}_{t^{*}_{m}}|Z^{(i)}_{t_{1}},\ldots,Z^{(i)}_{t_{n}} = g^{-1}_{t^{*}_{1}}(W^{(i)}_{t^{*}_{1}}),\ldots,g^{-1}_{t^{*}_{m}}(W^{(i)}_{t^{*}_{m}})|g^{-1}_{t_{1}}(W^{(i)}_{t_{1}}),\ldots,g^{-1}_{t_{n}}(W^{(i)}_{t_{n}}), $$ where  $g^{-1}_{t}(x) = \frac{\log(x)}{t}, ~t \neq 0\in \mathcal{T}$, is multivariate Gaussian and can be computed analytically. Then by transforming back to the original space, we have that for  $i \in \{1,2\}$:
\begin{align*}
&W^{(i)}_{t^{*}_{1}},\ldots,W^{(i)}_{t^{*}_{m}} \,|\, W^{(i)}_{t_{1}}=w_{t_{1}},\ldots,W^{(i)}_{t_{n}} =w_{t_{n}} \\
&=g_{t^{*}_{1}} (Z^{(i)}_{t^{*}_{1}}),\ldots,g_{t^{*}_{m}}(Z^{(i)}_{t^{*}_{m}}) \,|\, g_{t_{1}}(Z^{(i)}_{t_{1}}) = w_{t_{1}},\ldots,g_{t_{n}}(Z^{(i)}_{t_{n}}) = w_{t_{n}} 
\end{align*}

Finally, by applying the formula for the conditional distribution of a mixture of distributions from Theorem \ref{thm:extension_mixture}, we get the conditional distribution of $W_{t}$ at the new locations.

\begin{figure}
    \centering
    \includegraphics[width=0.5\linewidth]{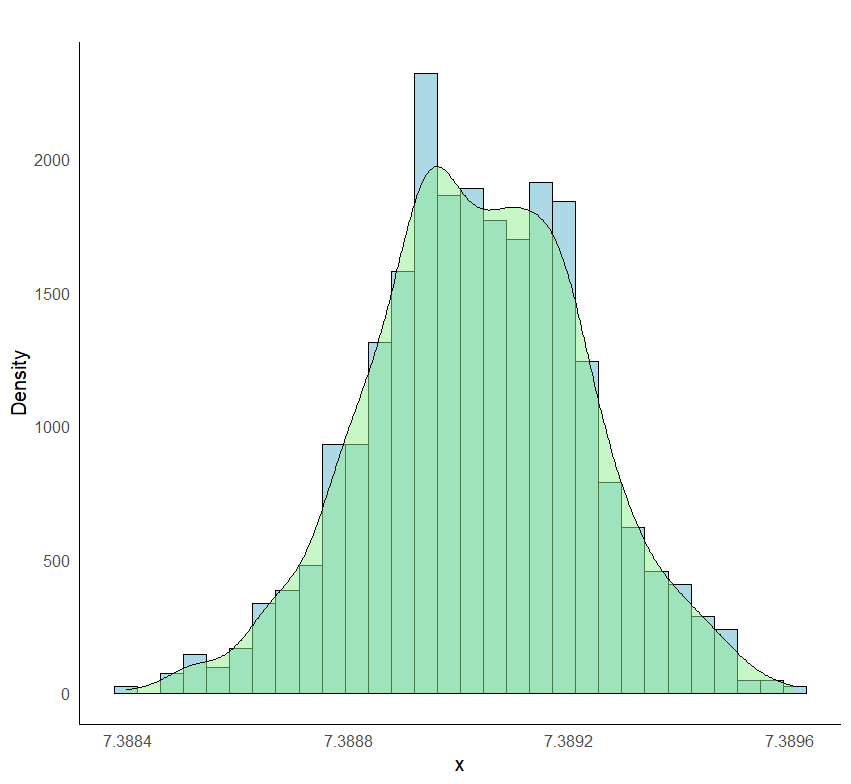}
    \caption{Conditional density of the transformed Gaussian Mixture process at location $t^{*}=0.5$}
    \label{fig:conditional_process}
\end{figure}

We provide in Figure \ref{fig:conditional_process} a numerical simulation showing an obtained conditional density with such a mixture of transformed Gaussian processes, when looking at a single new location $t^{*}= 0.5$ when having observed points $\mathbf{t} = (t_{1},\ldots,t_{n})$ on a grid. Additional details about this experiment can be found in Appendix \ref{sec:experiment_details}.

This is a specific instance of a copula process as defined by Wilson and al. in \cite{wilson2010copula}. However, formally extending the concept of stability under conditioning to copula processes involves complex technicalities, including the use of disintegrations, which falls beyond the scope of this paper.

\subsection{Probabilistic imputation}
\label{sec:imputation}
One key advantage of the proposed approach is its flexibility: once the joint distribution is learned, we can analytically compute the conditional distribution for any combination of observed and missing variables %(subject only to the calculation of the quantile function which we will discuss below).
This property is particularly valuable for imputation tasks, where different variables may be missing for different observations. In contrast, alternative distributional regression methods typically require fitting a separate model for each pattern of missing and observed variables. 
%—an approach that quickly becomes impractical as dimensionality increases.

%Our framework proceeds as follows. 
Starting with a complete dataset, we randomly introduce missingness in a given proportion of entries (here, $10\%$). Observations with no missing values form the training set, while those with one or more missing values comprise the test set. We fit a Gaussian Mixture Copula Model (GMCM) to the training data. For each test observation, we sample from the conditional distribution of its missing variables, conditioned on its observed values. This produces multiple imputed datasets.

\vspace{1em}

To evaluate imputation quality, we employ scoring rules. At first glance, applying scoring rules appears challenging: typically, one calculates an average score across all test predictions, but in our scenario, predictions may have different dimensions, complicating fair comparison.

One approach to circumvent this issue is to "complete" all predictions to joint distributions by plugging in the true observed values. However, it is worth questioning if evaluating these completed joint samples aligns with our actual imputation goal. Notably, for the Energy Score, it turns out that evaluating either the joint samples or the conditional samples of varying dimension is equivalent. In particular, for univariate predictions, this process reduces to calculating the CRPS.

\vspace{1em}
%At the end of our method, we need to convert the conditional samples from the latent space back to the original space. Traditionally, this involves using the quantile function of a Gaussian Mixture Model (GMM). However, as discussed in Section \ref{sec:fitting_gmcm}, calculating the quantile function of a GMM is complex and often requires costly bisection or optimization methods. In this section, we introduce a novel approach that eliminates the need to evaluate the quantile function of a GMM. Instead, we utilize the quantiles of truncated Gaussian distributions, which can be easily derived from the quantiles of a standard Gaussian distribution. The detail of this sampling algorithm is presented in the appendix.

We compared our approach with MICE \cite{van2011mice}, a widely used multiple-imputation procedure using chained equations. Our method was applied to a subset of four variables from the Breast Cancer Wisconsin dataset, see section \ref{sec::wisconsin}, known for its multimodal structure, as well as a simulated dataset generated from a GMCM. For further details, we refer to Appendix \ref{sec:imputation_experiment_details}.  %We applied our method on the Breast Cancer data Wisconsin \ref{sec::wisconsin} and a simulated dataset generated from a GMCM.

The results in Tables \ref{tab:imputation_simulated} and \ref{tab:imputation_wisconsin} indicate that GMCM slightly outperforms MICE in terms of Energy Score while requiring significantly less computational time. In experiments with a small number of features, our method performs comparably to MICE yet consistently shows a clear computational advantage. However, as the dimensionality increases, MICE tends to perform better. These findings suggest that our algorithm is well-suited for imputation tasks in datasets with few features and a multimodal structure, providing competitive results to MICE at a reduced computational cost.

\begin{table}[H]
    \centering
    \caption{Imputation results for the Wisconsin data set}
    \label{tab:imputation_wisconsin}
    \begin{tabular}{lcc}
        \toprule
        Metric                  & GMCM   & MICE    \\
        \midrule
        Energy Score            & \textbf{0.252}  & 0.286   \\
        Computational Time (s)  & \textbf{102.1} & 180.7   \\
        \bottomrule
    \end{tabular}
\end{table}

\vspace{1em}

\begin{table}[H]
    \centering
    \caption{Imputation results for simulated data}
    \label{tab:imputation_simulated}
    \begin{tabular}{lcc}
        \toprule
        Metric                  & GMCM   & MICE    \\
        \midrule
        Energy Score            & \textbf{0.097}  & 0.102   \\
        Computational Time (s)  & \textbf{158.3}  & 230.03   \\
        \bottomrule
    \end{tabular}
\end{table}

\section{Conclusion and perspectives}
\label{sec:conclusion}
In this work, we explored families of multivariate distributions that allow for analytical computation of conditional distributions. We provided examples beyond the Gaussian case, such as the multivariate Student $t$-distribution or the multivariate skew normal distribution.

Additionally, we introduced the concept of trans-dimensional families of multivariate distributions that are stable by conditioning. We demonstrated that this class extends to finite mixtures and increasing transformations, showing that a broader range of multivariate distributions can benefit from analytical conditioning. Also, we highlighted that stochastic process constructed through mixtures and transformations of processes with easy to condition finite-dimensional distributions remain easy to condition

Leveraging these results, we developed a versatile approach %generative algorithm 
for estimating conditional distributions from data. In practice, we model joint distributions using meta distributions with copulas that are stable by conditioning. In these settings, conditioning in original space is performed via a latent space where conditioning is more straightforward.

While our method allows for a high flexibility when considering broader families enjoying stability by conditioning, we focused in applications on the Gaussian Mixture Copula Model (GMCM) for practical purposes. In turn, we empirically compared different approaches to tackle the associated copula parameter fitting problem in Appendix \ref{sec:gmcm}.

Numerical results indicate that our approach can accurately estimate conditional distributions in simple synthetic cases. Although distributional random forests offer competitive performances in  estimating conditional distributions while being substantially faster, they do not enjoy some of the key advantages enabled by easy to condition probabilistic models. First, they provide (semi-) analytical forms for conditional distributions, facilitating post-hoc computations of various quantities such as conditional moments, probabilities, and other functionals.  Our method also supports the conditioning of stochastic processes, maintains coherence across conditioning queries, and allows modeling tail events. Furthermore, their generative nature is valuable for tasks like synthetic data generation and enables amortized inference for any conditional distribution, making it especially useful for imputation tasks, notably when missing data patterns are not known in advance and may evolve over time. Finally, our approach can handle conditioning with respect to linear combinations of components, which is particularly valuable for key tasks in uncertainty quantification, such as linear inverse problems \cite{travelletti2023uncertainty}. % Experiences on real datasets shows that the performance of our method performs well on several datasets, especially in cases where the data exhibits a multi-modal nature. Moreover, our method offers several advantages over competitors: it is interpretable, provides an analytical form for the conditional distribution, and enables amortized inference for any combination of missing variables, which motivates its application to imputation tasks. 

Given the promising results of our method in the considered continuous framework, an important next step is to extend this methodology to handle mixed variables, which can include continuous and/or discrete components. Such an extension would be particularly relevant in medical applications where health outcomes of interest are often measured as categorical or dichotomous states. A possible route would be to assume the discrete variables are discretizations of latent continuous variables as is common practice in latent class modeling frameworks.
One limitation of our work is the challenge of fitting copulas, especially as dimensionality increases. Specifically, for the GMCM, the number of parameters scales quadratically with dimension, making it feasible only for relatively low-dimensional problems. Moreover, we currently need to manually preset the number of mixture components in the GMCM, whereas an automated method for determining this would be beneficial.

In higher-dimensional settings, other generative methods for sampling from a conditional density use tensor-train decomposition to approximate the joint density and sample from the conditional distribution by using the inverse Rosenblatt transform \cite{cui2023scalable}. This approach has been shown to have strong theoretical guarantees, however it requires a predetermined ordering of variables, which can limit its usefulness in tasks like imputation.
%transform the discrete variables to a latent continuous space, apply our method, and then transform back to the space of interest.

Future research directions include extending our method by implementing the algorithm for other dependence structures, such as mixtures of Student t-distributions and mixtures of skew normal distributions. Additionally, we aim to extend our work to accommodate missing values in the observed sample.

\section*{Acknowledgments.} The authors gratefully acknowledge the Isaac Newton Institute for its support during the program "Representing, Calibrating, $\&$ Leveraging Prediction Uncertainty from Statistics to Machine Learning," where part of this work was conducted. Additionally, we extend our thanks to Dr. Jan Vanhove for his valuable implementation suggestions, Prof. Dr. Marina Eckermann for directing us to the dataset on multiscale imaging of the human hippocampus, and Prof. Dr. Wolfgang Polonik for his insightful remarks.

\bibliographystyle{siamplain}
\bibliography{references}

\appendix
\section{Infinite mixtures}
\label{sec:infinite}

Let us consider a mixing measure $\pi$ such that the mixture distribution can be written:
    $$f(x_{1},\ldots,x_{d}) = \int f(x_{1},\ldots,x_{d}|\psi) d\pi(\psi), ~\forall (x_{1},\ldots,x_{d})$$

For every pair of strictly ordered indices $(\mathbf{i}, \mathbf{j}) \in  I_{\ell,m}^{d}$ and conditioning variables 

$(x_{j_{1}},\ldots,x_{j_{m}}) \in \prod_{k=1}^{m} \mathcal{X}_{j_{k}}$, we have that $\forall (x_{i_{1}},\ldots,x_{i_{\ell}}) \in \prod_{k=1}^{\ell} \mathcal{X}_{j_{k}}$:
\begin{equation*}
\begin{split}
 f_{\mathbf{i}|\mathbf{j}} (( x_{i_{1}},\ldots, x_{i_{\ell}})|(x_{j_{1}},\ldots,x_{j_{m}})) &= \frac{f_{[\mathbf{i},\mathbf{j}]}( x_{i_{1}},\ldots, x_{i_{\ell}},x_{j_{1}},\ldots,x_{j_{m}})}{f_{\mathbf{j}}(x_{j_{1}},\ldots,x_{j_{m}})} \\
 &=  \int \frac{f_{[\mathbf{i},\mathbf{j}]}( x_{i_{1}},\ldots, x_{i_{\ell}},x_{j_{1}},\ldots,x_{j_{m}}|\psi)}{f_{\mathbf{j}}(x_{j_{1}},\ldots,x_{j_{m}})} d\pi(\psi)   \\
 &= \int  \frac{f_{[\mathbf{i},\mathbf{j}]}( x_{i_{1}},\ldots, x_{i_{\ell}},x_{j_{1}},\ldots,x_{j_{m}}|\psi)}{f_{\mathbf{j}}(x_{j_{1}},\ldots,x_{j_{m}}|\psi)} \frac{f_{\mathbf{j}}(x_{j_{1}},\ldots,x_{j_{m}}|\psi)}{f_{\mathbf{j}}(x_{j_{1}},\ldots,x_{j_{m}})} d\pi(\psi)\\
 &= \int  f_{\mathbf{i}|\mathbf{j}} (( x_{i_{1}},\ldots, x_{i_{\ell}})|(x_{j_{1}},\ldots,x_{j_{m}}, \psi)) \frac{f_{\mathbf{j}}(x_{j_{1}},\ldots,x_{j_{m}}|\psi)}{f_{\mathbf{j}}(x_{j_{1}},\ldots,x_{j_{m}})} d\pi(\psi) ,
\end{split}
\end{equation*}
The conditional distribution of the mixture can be written as an infinite mixture of the individual conditional distributions with mixing measure $\Tilde{\pi} = \frac{f_{\mathbf{j}}(x_{j_{1}},\ldots,x_{j_{m}}|\psi)}{f_{\mathbf{j}}(x_{j_{1}},\ldots,x_{j_{m}})} d\pi(\cdot)$.

\vspace{1em}

\section{Multivariate Student $t$-distribution}

\begin{definition}(Multivariate Student $t$-distribution)

\label{def::student}
      If $\textbf{Y} \sim \mathcal{N}_{d}(0,\Sigma)$ and $u\sim \chi^{2}_{\nu}$  then $\frac{\textbf{Y}}{\sqrt{\frac{u}{\nu}}} = \textbf{X} - \mu$ with $\textbf{X} \sim t_{d}(\mu,\Sigma,\nu)$, where $\chi^{2}_{\nu}$ represents a chi-square distribution with $\nu$ degrees of freedom. We say that $\textbf{X}$ follows a n-multivariate t-distribution with mean $\mu$, covariance matrix $\Sigma$ and $\nu$ degrees of freedom.
\end{definition}

The multivariate Student $t$-distribution is stable by marginalization and conditioning as states the two following propositions (\cite{ding2016conditional}). In this appendix, we will use fixed, ordered sets of indices, with the first set containing the variables we wish to condition on the variables in the second set. These results can be readily extended to any arbitrary set of indices.

\begin{proposition}(Marginal distribution)
\label{prop::marginal_student}
    Let $\textbf{X} = (\mathbf{X_{1}}, \mathbf{X_{2}})^{t} \sim t_{d}(\mu,\Sigma,\nu)$ with $\mu = (\mu_{1}, \mu_{2})^{t}$, $\mu_{1} \in \mathbb{R}^{\ell}$ , $\mu_{2}\in \mathbb{R}^{m}$, $\ell+m=d$ and \[\Sigma =\begin{pmatrix}
            \Sigma_{1,1} & \Sigma_{1,2} \\
            \Sigma_{2,1} & \Sigma_{2,2} 
            \end{pmatrix}\]
     then the marginal distribution of $\mathbf{X_{2}}$ is given by:
     $$\mathbf{X}_{2} \sim t_{m}(\mu_{2},\Sigma_{22},\nu)$$
            
\end{proposition}

The marginal distribution is again a multivariate Student t-distribution.
\begin{proposition}(Conditional distribution)

\label{prop::cond_student}
    With the same notation as in the proposition above, the conditional distribution of $\mathbf{X}_{1}$ given that $\mathbf{X}_{2} = \mathbf{x_{2}}$ is:
    $$\mathbf{X_{1}}|\mathbf{X}_{2} = \mathbf{x_{2}} \sim t_{\ell}\left (\mu_{1.2},\frac{\nu + d_{2}}{\nu + \ell}\Sigma_{11.2},\nu+\ell\right)$$ with $\mu_{1.2}=\mu_{1} + \Sigma_{12} \Sigma_{22}^{-1}(\textbf{x}_{2}-\mu_{2})$ , $\Sigma_{11.2}=\Sigma_{11}-\Sigma_{12} \Sigma_{22}^{-1} \Sigma_{21}$ and $d_{2} =(\textbf{X}_{2}-\mu)^{t}\Sigma_{22}^{-1} (\textbf{X}_{2}-\mu) $.
\end{proposition}
The conditional distribution is again a multivariate Student t-distribution but with another degrees of freedom. We say that the characteristic generator has been modified through conditioning. 

\section{Multivariate unified skew distributions}
Another class of multivariate distributions which are stable by conditioning and marginalization  property are the multivariate unified skew normal and skew Student $t$-distributions. Unlike the Gaussian distribution, they allow for skewness.

Multivariate skew normal distributions can be defined in different ways by extending the univariate skew-normal density to the $d$-dimensional
case (\cite{azzalini1996multivariate}). Here we are mainly interested in the \textit{Multivariate Unified Skew Normal} distribution, we will explain his stochastic construction and conditioning properties.

Let $U=\begin{bmatrix}\mathbf{U}_{0}\\\mathbf{U}_{1}\end{bmatrix} \sim \mathcal{N}_{p+d}(0,\Omega^{*})$ with $\Omega^{*} = \begin{bmatrix}\Gamma  & \Delta^{T}\\\Delta & \Bar{\Omega}\end{bmatrix}$, $\mathbf{Z} = (\mathbf{U_{1}}| \mathbf{U_{0}} + \gamma >0)$ , with $\gamma \in \mathbb{R}^{d}$,
and let define $\mathbf{Y} = \mathbf{\xi} + \mathbf{\omega} \mathbf{Z} $ with $\mathbf{\xi} \in \mathbb{R}^{d}$ and $\omega =\text{diag}(\omega_{1},\ldots,\omega_{d}) >0$.

The density function of a variable of type $(\mathbf{X}_{1} | \mathbf{X}_{0} > c)$, evaluated at point $\mathbf{x}$, is most easily
computed via the general relationship

$$f(\mathbf{x}) = \frac{f_{X_{1}}(\mathbf{x}) \mathbb{P}(X_{0}>c|X_{1} = \mathbf{x})}{ \mathbb{P}(X_{0}>c)}$$
It can be shown that the
density of $\mathbf{Y} $ at $\mathbf{x} \in  \mathbb{R}^{d}$
is:
\begin{equation*}
    \label{eq::dens_skew_normal}
    f_{Y}(\mathbf{x}) = \phi_{d} (\mathbf{x}-\xi;\Omega) \frac{\Phi_{p}(\gamma + \Delta^{T} \Bar{\Omega}^{-1}\omega^{-1}(\mathbf{y}-\xi));\Gamma- \Delta^{T} \Bar{\Omega}^{-1} \Delta)}{\Phi_{p}(\gamma;\Delta)}
\end{equation*}
then we write $Y \sim SUN_{d,p}(\xi,\gamma,\Bar{\omega},\Omega^{*})$
with $\Bar{\omega}=\omega 1_{d}$ the vector of diagonal elements of $\omega$, and we say that $Y$ as a multivariate skew unified normal distribution (\cite{arellano2006unification}).

The two following properties shows that the multivariate skew unified normal distribution is closed under marginalization and conditioning.

\begin{proposition}(Marginal distribution)

\label{prop::marg_skew_normal}
Let $\mathbf{X} = (\mathbf{X_{1}},\mathbf{X_{2}})^{t} \sim SUN_{d,p}(\xi,\gamma,\Bar{\omega},\Omega^{*})$ with $\mathbf{X_{1}} \in \mathbb{R}^{\ell}$ and $\mathbf{X_{2}} \in \mathbb{R}^{m}$ with $\ell +m =d$, $\xi = (\xi_{1},\xi_{2})^{t} $ with $\xi_{1} \in \mathbb{R}^{\ell}, ~\xi_{2} \in \mathbb{R}^{m}$ and $\bar{\omega}_{2} = \text{diag} (\omega_{\ell+1},\ldots,\omega_{d}) \in \mathbb{R}^{m \times m}$ .
The marginal distribution of $\mathbf{X_{2}}$ is
$$\mathbf{X}_{2} \sim SUN_{m,p}(\xi_{2}, \gamma,\Bar{\omega}_{2},\Omega^{*}_{2})$$ where $\Omega^{*}_{2} = \begin{bmatrix}\Gamma & \Delta_{2}^{t}\\\Delta_{2} & \Bar{\Omega}_{22}\end{bmatrix} $.
    
\end{proposition}

\begin{proposition}(Conditional distribution)

\label{prop::cond_skew_normal}
    With the same notations as above, the conditional distribution of $\mathbf{X}_{1}$ given $\mathbf{X}_{2} =\mathbf{x}_{2}$ is given by
$$(\mathbf{X}_{1}|\mathbf{X}_{2}=\mathbf{x}_{2}) \sim SUN_{\ell,p}(\xi_{1.2},\gamma_{1.2},\Bar{\omega}_{1},\Omega^{*}_{22.1})$$
where $\xi_{1.2} = \xi_{1} + \Omega_{12} \Omega_{22}^{-1}(\mathbf{y}_{2}-\xi_{2})$, $\gamma_{1.2} = \gamma + \Delta_{2}^{T} \Bar{\Omega}_{22}^{-1} \omega_{2}^{-1}- (\mathbf{y}_{2}-\xi_{2})$
and
$$\Omega^{*}_{22.1} = \begin{bmatrix}\Gamma_{1.2}  & \Delta_{1.2}^{T}\\\Delta_{1.2} & \Bar{\Omega}_{11.2}\end{bmatrix}$$
with 
$\Omega^{*}_{11.2} = \Bar{\Omega_{11}} - \Bar{\Omega}_{12} \Bar{\Omega}_{22}^{-1} \Bar{\Omega}_{21} $,
$\Delta_{1.2} = \Delta_{1} - \Bar{\Omega}_{12}\Bar{\Omega}_{22}^{-1} \Delta_{2}$ and $\Gamma_{1.2} = \Gamma - \Delta_{2}^{T} \Bar{\Omega}_{22}^{-1}\Delta_{2}$.

\end{proposition}
With the stochastic representation of a multivariate skew unified normal distribution detailed above, we can generate samples from this distribution if we know how to sample from a multivariate normal distribution.

\section{Details on the GMCM}
\label{sec:details_gmcm}
\subsection{Fitting of a GMCM}
\label{sec:fitting_gmcm}
%In order to fit Gaussian mixture copula models, the \textit{R} package \textit{GMCM} is commonly used (\cite{bilgrau2016gmcm}) but some alternatives in \textit{Python} using the package \textit{Tensorflow} have been developed lately.
Estimating parameters of a GMCM is a challenging task for several reasons as it has been discussed in \cite{tewari2023estimation}.

The first challenge is that 
copula invariance under marginal increasing transformations causes %copulas are invariant under marginal increasing transformations. Specifically, if $\mathbf{X} = (X_{1},\ldots,X_{d})^{t}$ is a random vector with continuous marginal distributions and copula $C$, and $T_{1},\ldots,T_{d}$ are strictly increasing functions. Then $(T_{1}(X_{1}),\ldots,T_{d}(X_{d}))^{t}$ also has copula $C$.
%In the case of a GMCM, this property implies 
a lack of identifiability of the parameters. 
%which makes the fitting of the parameters more difficult. 
This can be seen through the following proposition \cite{tewari2023estimation}:

\begin{proposition}(Non identifiability of GMCM)

    Let $\theta = \{ \alpha_{i},\mu_{i}, \Sigma_{i}\}_{i=1}^{K}$ and $\theta'= \{ \alpha_{i},A \mu_{i} + \mathbf{b}, A^{t} \Sigma_{i} A\}_{i=1}^{K}$ two valid set of parameters with $\mathbf{b} \in \mathbb{R}^{d}, ~A \in \mathbb{R}^{d\times d}$ diagonal positive definite matrix. Then the likelihood of the two Gaussian mixture copula models with these parameters are the same i.e
    $$\forall \mathbf{u} \in \mathbb{R}^{d}, ~c_{gmc}(u_{1},\ldots,u_{d};\theta) = c_{gmc}(u_{1},\ldots,u_{d};\theta').$$
\end{proposition}

This lack of identifiability results in the likelihood of a GMCM having infinitely many maximizers. To address this issue,  Bilgrau \cite{bilgrau2016gmcm} proposed standardizing the first component to have zero mean and unit variance along each dimension. 
%which resolves the identifiability issue in most cases. 

Another challenge in fitting GMCM is the high computational cost associated with functional evaluations, as the quantile function of a Gaussian mixture model isn't known in closed form. To address this, we can use numerical approximation schemes leveraging the monotonicity of the cdf, although they tend to be computationally expensive. Alternatively, methods based on the bisection algorithm, such as the one developed by Chandrupatla \cite{chandrupatla1997new}, can be employed to find the inverse of a cdf. Since the derivatives of the Gaussian mixture model’s cdf are known analytically, a Newton-type algorithm can also be effectively used in this context.

In the literature, two primary methods have been proposed for estimating the parameters of a GMCM via direct gradient-based optimization. The classical approach involves using finite difference methods (denoted as FD in the rest of this paper) to estimate the derivatives. Approximating derivatives numerically can cause imprecisions and heavy computational costs.

Recently, a more efficient method, leveraging modern probabilistic programming language tools such as automatic differentiation (denoted as AD), has been developed. Automatic differentiation computes the exact values of the derivatives by leveraging the chain rule, significantly enhancing both computational efficiency and accuracy.

Another possible approach for parameter estimation is using an Expectation-Maximization (EM) type algorithm. However, due to the changing inverse distribution functions along the margins with every parameter update, the inputs of the EM algorithm are not fixed, rendering the traditional EM algorithm unsuitable. Although a pseudo-EM algorithm (denoted as PEM) has been proposed and discussed in the literature, it lacks convergence guarantees and often shows suboptimal performance.

The methods FD and PEM are implemented in the \textit{R} package \textit{GMCM} \cite{bilgrau2016gmcm}.Meanwhile, the AD method has been developed in \textit{Python} by Tewari in a Github page related to this paper \cite{tewari2023estimation}.

\subsection{Comparison of the copula fitting methods}
To compare the different optimization methods described above (AD, FD and PEM), we simulate data following a GMCM with known parameters and attempted to recover its values using different fitting procedures. Because GMCM parameters are not identifiable, we cannot directly compare the recovered parameter values. Instead, we chose to compare metrics such as the log-likelihood or distances to the true distribution. We used the Cramér distance to quantify the differences between empirical multivariate distributions. We compared samples of the true underlying distributions to samples of the distribution with the fitted parameters and repeated the experiments several times to take into account the uncertainty in the sampling of a GMCM. In the results, we indicated the average Cramér distance and average log-likelihood as well as the standard deviations.

To ensure generality, we simulated data from several parameter configurations with varying dimensions and numbers of components, see details in the appendix.

\begin{center}
\begin{table}[H]%
\caption{Fitting performances of the different optimization methods for a $2D$ example in terms of distance to a sample from the true underlying distribution.}
  \vspace{1em}
  \label{table:cut-off_0.01} 
\begin{tabular*}{400pt}{@{\extracolsep\fill}lcccccc@{\extracolsep\fill}}
 & \multicolumn{2}{c}{\textbf{Log-likelihood}} & \multicolumn{2}{c}{\textbf{Cramer}} \\  
 & \textbf{Mean} & \textbf{SD} & \textbf{Mean} & \textbf{SD} \\  
 \toprule 
\textbf{AD} & 208.7 & 16.9 & 0.278 &  $8.4 \times 10^{-5}$ \\ 
\textbf{PEM} & -151.7 & 34.2 & 0.422 & $9.1 \times 10^{-5}$ \\ 
\textbf{FD} & -154.4 & 35.3&  0.406 & $8.8 \times 10^{-5}$ \\
\bottomrule
\end{tabular*}
\end{table}
\end{center}

\begin{center}
\begin{table}[H]%
\caption{Fitting performances of the different optimization methods for a $3D$ example in terms of distance to a sample from the true underlying distribution.}
  \vspace{1em}
  \label{table:cut-off_0.01} 
\begin{tabular*}{400pt}{@{\extracolsep\fill}lcccccc@{\extracolsep\fill}}
 & \multicolumn{2}{c}{\textbf{Log-likelihood}} & \multicolumn{2}{c}{\textbf{Cramer}} \\  
 & \textbf{Mean} & \textbf{SD} & \textbf{Mean} & \textbf{SD} \\  
 \toprule 
\textbf{AD} & 1104.2 & 44.1 & 0.330 & $8.9 \times 10 ^{-5}$\\ 
\textbf{PEM} & 1069.2 & 35.7 & 0.360 & $8.80 \times 10 ^{-5}$ \\ 
\textbf{FD} & 1063.8 & 43.7&  0.364 & $9.3 \times 10 ^{-5} $ \\
\bottomrule
\end{tabular*}
\end{table}
\end{center}

From the results above, it seems that the direct optimization of the likelihood function by using automatic differentiation tools (AD) performs better than the other methods for several parameter values.
That is why we rely on this method for GMCM fitting in the experiments presented next.

\section{Experiment details}
\label{sec:experiment_details}

\subsection{GMCM fitting}

For the AD implementation, we utilized the Adam optimizer with a learning rate of \(10^{-3}\). For the FD method, the Nelder-Mead optimizer was employed, with a maximum of 10,000 iterations. Similarly, the PEM method also used a maximum of 10,000 iterations.
The true values of the parameters are the following:
\begin{itemize}
    \item \textbf{2D example:}

    $\alpha_{1} = 0.45, ~\alpha_{2} = 0.55$, 
$\mu_{1} = \begin{pmatrix}
    5.15 \\
    4.32
\end{pmatrix}, ~\mu_{2} =  \begin{pmatrix}
    -20.07 \\
    3.04
\end{pmatrix}$, $\Sigma_{1} = \begin{pmatrix}
    5.6 & 2.3 \\
    2.3 & 8.27
\end{pmatrix}, ~ \Sigma_{2} = \begin{pmatrix}
    3.35 & 1 \\
    1 & 1.16
\end{pmatrix}$

\vspace{1em}

     \item \textbf{3D example:}

    $\alpha_{1} = 0.69, ~\alpha_{2} = 0.163,~\alpha_{2} = 0.147$, 
$\mu_{1} = \begin{pmatrix}
    1.19 \\
    5.63 \\
    -9.67
\end{pmatrix}, ~\mu_{2} =  \begin{pmatrix}
    -6.75 \\
    12.04 \\
    -8.44
\end{pmatrix}, ~\mu_{3} =  \begin{pmatrix}
    -5.92 \\
    -4.0 \\
    2.58
\end{pmatrix}$,  $\Sigma_{1} = \begin{pmatrix}
    2.26 & 1.33 & -1.163544 \\
    1.33 &  2.71 &-1.36 \\
    -1.16 & -1.36 & 3.78

\end{pmatrix}, ~ \Sigma_{2} = \begin{pmatrix}
   12.24 & 4.14 & -3.87 \\
   4.14 & 10.33 &  2.76 \\
   -3.87 & 2.76 & 16.45
\end{pmatrix}$,

$
 \Sigma_{3} = \begin{pmatrix}
   3.19 & -0.38  & 0.54 \\
   -0.38 & 0.76 &-0.02\\
   0.54 & -0.02   & 1.01
\end{pmatrix}$
\end{itemize}

%\subsection{CKDE Benchmark}

%The CKDE method used as benchmark, fits a Kernel Density Estimation (KDE) model (using a Gaussian kernel and a bandwidth of 1 by using the function \textit{KernelDensity} of the package \textit{sklearn} in \textit{Python}) on the joint distribution of the response variables and predictors $\mathbf{X}=(\mathbf{X_{1}},\mathbf{X_{2}})$ . It then generates samples from the conditional distribution \(\mathbb{P}(\mathbf{X_{1}}|\mathbf{X_{2}})\). This is achieved by drawing an oversampled set of joint samples from the KDE model, filtering those to include only the samples close to a given predictor value \(\mathbf{X}_{\text{cond}}\) within a specified tolerance (chosen to be 0.4 for most of the experiments, except for the experiment on the Wine dataset where it is chosen to be 10), and finally selecting the required number of samples (1000) from this filtered set. This process allows obtaining samples from the conditional distribution without needing an analytical form of the joint or conditional distributions.

\subsection{Distributional random forests benchmark}

As a benchmark, we considered the distributional random forest (DRF) method
developped by Cevid and al.  \cite{cevid2022distributional}. We used the implementation of the package \textit{drf} from the  \textit{R} software.

\subsection{Stochastic process experiment.}

We generated two Gaussian processes $Z^{(1)}_{t}$ and $Z^{(2)}_{t}$ on a grid of $100$ uniformly spaced points on $[-3,3]$, denoted $t_{1},\ldots,t_{100}$. We consider $Z^{(1)}_{t}$ to have a mean of $0$ and a Gaussian covariance kernel with bandwidth $\sigma = 0.5$ and  $Z^{(2)}_{t}$ to have a mean of $1$ and a Gaussian covariance kernel with bandwidth $\sigma = 0.5$. As described in Section \ref{sec:cond_process}, we consider the transformations $g_{t} : x \rightarrow \exp{(tx)}$ and $p=0.6$. We sample from the conditional distribution at the new location $t^{*} = 2$, conditioned on $Z^{(i)}_{t_{1}}, \ldots, Z^{(i)}_{t_{100}} = 1$ for $i \in \{1,2\}$.
\subsection{Imputation experiment}
\label{sec:imputation_experiment_details}
The dataset for the imputation experiment on simulated data \ref{sec:imputation} has been generated by sampling $1000$ points from a Gaussian Mixture Copula  Model (GMCM) with two components, with means $\mu_{1} = (1,2)^t$ and $\mu_{2}=(3,4)^{t}$, weights $w=(0.6,0.4)$ and covariances matrices:
$$\Sigma_1 =\begin{pmatrix}
            1 & 0.5 \\
            0.5 & 1 
            \end{pmatrix},~~ \Sigma_2 =\begin{pmatrix}
            1 & 0.5 \\
            0.5 & 1 
            \end{pmatrix}$$
The marginal distributions have been chosen to be from a GMM with respectively means $\mu_{m,1} = (1,2)^t $ and $\mu_{2}=(3,4)^{t}$, weights $w_{1} = (0,4,0.6)$ and $w_{2} = (0.4,0.6)$, and standard deviations $\sigma_{1}= (1,1)^{t}$ and $\sigma_{2} = (1,1)^{t}$.

For the experiment with the Breast Cancer Wisconsin data, we focused on a subset of four variables: perimeter standard error (PSE), extreme smoothness (ES), extreme concavity (EC), and extreme concave points (ECP).

In both these experiments, we used the function \textit{qmixnorm} from the \textit{R} package \textit{KScorrect} to compute quantiles of a Gaussian Mixture Model.

The computer used for this experiment is a HP Elite Book 845 with a a AMD Ryzen 7 6800U processor and 16 Go of RAM.
\subsection{Scoring rules}
The CRPS, Energy Score and Variogram Score have been implemented using directly the \textit{Scoringrules} package in \textit{Python}. To implement the logarithmic score, we estimated the density of the predicted values by kernel density estimation (with a Gaussian kernel and a bandwidth of 0.5) and then evaluate the log-likelihood of the observed value.
\end{document}